\documentclass[preprint,authoryear,12pt]{elsarticle}
 \usepackage{amssymb}
 \usepackage{amsthm}
 \usepackage{amsmath}
 \usepackage{amsmath,amssymb,amsthm,graphicx,caption,bm}
 \usepackage{amsfonts}
 \usepackage{mathrsfs}
 \usepackage{verbatim}
 \usepackage[usenames, dvipsnames]{color}
 \usepackage[capposition=below]{floatrow}
 \usepackage[section]{placeins}
 \usepackage[utf8]{inputenc}
 \usepackage{rotating}
 \usepackage[figurename=Figure]{caption}
 \usepackage[colorlinks=true,linkcolor=blue]{hyperref}%
 \usepackage{lscape}
 \usepackage{lineno}

  \newcommand{\Lower}[1]{\smash{\lower 1.5ex \hbox{#1}}}

  \def\vep{\varepsilon}
  
  \def\R{\mathbb{R}}
  
  \def\bX{\bm{X}}
  
  \def\bY{\bm{Y}}
  \def\bW{\bm{W}}

  \def\bZ{\bm{Z}}
  \def\bT{\bm{T}}
  
  \def\bbt{\boldsymbol{\beta}}

  \def\argmin{\arg\hskip -0.02in\min}
  
  \newenvironment{eqarray*}{\arraycolsep 0.14em\begin{eqnarray*}}{\end{eqnarray*}}
  \newtheorem{thm}{Theorem}

 \numberwithin{equation}{section}
 \journal{Submitted to TBA}

\begin{document}

\begin{frontmatter}

\title{{Extrapolation Estimation for Parametric Regression with Normal Measurement Error}}
\author{Kanwal Ayub}
\author{Weixing Song\corref{cor}}
\ead{weixing@ksu.edu}
\cortext[cor]{Corresponding author}
\address{Department of Statistics, Kansas State University, Manhattan, KS 66506}

\begin{abstract}
  For the general parametric regression models with covariates contaminated with normal measurement errors, this paper proposes an accelerated version of the classical simulation extrapolation algorithm to estimate the unknown parameters in the regression function. By applying the conditional expectation directly to the target function, the proposed algorithm successfully removes the simulation step, by generating an estimation equation either for immediate use or for extrapolating, thus significantly reducing the computational time. Large sample properties of the resulting estimator, including the consistency and the asymptotic normality, are thoroughly discussed. Potential wide applications of the proposed estimation procedure are illustrated by examples, simulation studies, as well as a real data analysis.
\end{abstract}

 \begin{keyword} Parametric regression \sep Measurement Error \sep Simulation and Extrapolation \vskip 0.02in

 \MSC primary 62G05\sep secondary 62G08
\end{keyword}
\end{frontmatter}

\renewcommand\thelinenumber{\color{red}\arabic{linenumber}}


 \section{Introduction}

  As a simulation-based estimation method, the classical simulation-extrapolation (SIMEX) algorithm has enjoyed much popularity among its peers in the measurement error literature. Suppose a random variable $X$ is generated from a population whose distribution is characterized by an unknown parameter $\theta$.  Unfortunately, $X$ cannot be directly measured, instead, what we can observe is a surrogate variable $Z$. A commonly used structure in measurement literature assumes that $Z$ and $X$ are related through $Z=X+U$, where $U$ is called the measurement error, and is independent of $X$. We further assume that $U$ has a normal distribution with mean $0$ and variance $\sigma_u^2$ which is assumed to be known. If $X$ can be observed, suppose a statistic $T(\bX)$ can be found to estimate $\theta$ based on a sample $\bX=(X_1,\ldots, X_n)$ of size $n$ from $X$.

  Now suppose we have the data $\bZ=(Z_1,\ldots, Z_n)$.  A typical classical SIMEX procedure for estimating $\theta$ consists of the following three steps. In the first step, $n$ i.i.d. random numbers $V_i$'s are generated from $N(0,\sigma_u^2)$, and for a pre-specified nonnegative number $\lambda$, $\tilde Z_i(\lambda)=Z_i+\sqrt{\lambda}V_i$, $i=1,2,\ldots, n$ are calculated, and based on these pseudo-data, $T(\tilde\bZ(\lambda))$ is calculated, where $\tilde{\bZ}(\lambda)=\{\tilde Z_1(\lambda),\ldots, \tilde Z_n(\lambda)\}$. This operation is repeated for a large number of times, say $B$. Denote the resulting $T$-values as $T_b(\tilde\bZ(\lambda))$, $b=1,2,\ldots, B$. Finally, averaging these $B$ quantities concludes the first step, and we denote the resulting average by $\overline T(\lambda)$; In the second step, repeating the first step for a sequence of nonnegative $\lambda$ values, for example, $\lambda=\lambda_1,\ldots,\lambda_K$, we obtain $\overline T(\lambda_1),\ldots, \overline T(\lambda_K)$. In real applications, a rule-of-thumb for choosing the sequence of values $\lambda_1, \ldots, \lambda_K$ is to select $K$ equally spaced values from $[0,2]$ for $K$ around $20$. In the last step, a trend of $\overline T(\lambda)$ with respect to $\lambda$ is identified, the trend is then used to evaluate the value of $\overline T(\lambda)$ at $\lambda=-1$, and the extrapolated value $\overline T(-1)$ is the SIMEX estimate of $\theta$. For more information on the SIMEX procedure, see the seminal papers of \cite{cook1994} and \cite{stefan1995}. The asymptotic properties of the SIMEX procedure, when $\sigma^2$ is small, is investigated in \cite{carroll1996}. Although we briefly introduce the classical SIMEX procedure for the univariate $X$ cases, the multivariate scenarios are accommodated very well.

  In all the literature involving applications of the SIMEX procedure, the three steps described above are strictly followed. In particular, pseudo data are always generated to provide the values of $T(\tilde{\bZ}(\lambda))$, and an average follows. It is well known that the simulation step in the SIMEX procedure is notoriously time-consuming in the SIMEX procedure, in particular, when the estimator $T(\bX)$ has no closed-form and is determined inexplicit by an optimization process. The other drawback in using the classical SIMEX approach is the choice of the extrapolation function in the extrapolation step. Except for some very special cases, there are no tractable extrapolation functions to use in general. In fact, in most applications, the approximate ones, such as the linear, quadratic, and nonlinear forms, are taken as the working extrapolation functions. In this paper, we will construct an estimation procedure to avoid, or at least partially avoid, these drawbacks in the classical SIMEX procedure. Finally,
  it is also worth to mention that many papers indicate the classical SIMEX procedure is robust to the distributional assumption on the measurement error, that is, even when the measurement error is not normally distributed, the SIMEX procedure still provides reasonable estimates. However, \cite{song2014} theoretically proved that this is not true in general. Nevertheless, throughout this paper, we shall assume $U$ has a normal distribution.

  This paper is organized as follows. Some examples that motivate our research, such as the linear, quantile, and expectile regressions etc., are introduced in Section \ref{sec2}; the proposed extrapolation estimation procedure for the general parametric regression model is constructed in Section \ref{sec3}, together with an exploration on the large sample results. A discussion on the extrapolation function can also be found in this section; numerical studies are conducted in Section \ref{sec4}; Section \ref{sec5} includes a discussion on the potential extension of the proposed estimation procedure to some semi-parametric regression models, as well as the robustness of the proposed estimation procedure to the normality assumption. All the technical proofs of the theoretical results are deferred to the Appendix section.

 \section{Motivating Examples}\label{sec2}

  Before formulating our extrapolation estimation procedure for the general parametric regression models, we start with a very simple example to see how our research idea has been developed. Suppose we have a simple linear errors-in-variables regression model $E(Y|X)=\alpha+\beta X$, and $Z=X+U$. As discussed in \cite{carroll1999}, for any fixed $\lambda>0$, after repeatedly adding the extra normal measurement errors, and computing the ordinary least squares (LS) slope, the averaged estimator consistently estimates $g(\lambda)=\beta\sigma_x^2/(\sigma_x^2+(1+\lambda)\sigma_u^2)$, where $\sigma_x^2$ is the variance of $X$. Obviously, extrapolating $\lambda$ to $-1$ leads to $g(-1)=\beta$. This clearly shows that the SIMEX method works really well for the linear regression model. In fact, in the seminal paper of \cite{cook1994}, the SIMEX estimators of $\alpha$ and $\beta$ are derived without the simulation step. Instead, the conditional expectation of the least square estimates based on the pseudo-data given the observed data are calculated, and under the NON-IID pseudo-errors, the following estimators of $\alpha$ and $\beta$ are constructed:
     \begin{equation}\label{eq2.1}
       \hat\alpha(\lambda)=\overline Y-\hat \beta(\lambda)\overline Z,\quad
       \hat\beta(\lambda)=\frac{S_{YZ}}{S_{ZZ}+\lambda\sigma_u^2},
     \end{equation}
  where $\overline Y,\overline X$ are the sample means of $Y$ and $Z$, $S_{YZ}, S_{ZZ}$ are the sample covariance of $Y$ and $Z$, the sample variance of $Z$, respectively. Directly letting $\lambda=-1$ leads to the SIMEX estimator. For more details regarding the NON-IID pseudo-errors, please refer to \cite{cook1994} and Section 5.3.4.1 in \cite{carr2006}. In the following, we would like to show that the SIMEX estimators of $\alpha$, $\beta$ defined in (\ref{eq2.1}) can be obtained from another perspective, without using the so-called NON-IID pseudo-errors.

  Consider a multiple linear regression model and the LS procedure. When $X$ is observable, the LS estimators of $\alpha$ and $\beta$ can be estimated by minimizing the LS criterion
  $\sum_{i=1}^n(Y_i-\alpha-\beta^T X_i)^2$. Since $X_i$'s are not available, following the SIMEX idea, we generate the pseudo-data $Z_i(\lambda)=Z_i+\sqrt{\lambda}V_i$, $V_i\sim N(0,\Sigma_u)$, $i=1,2,\ldots,n$, where $\Sigma_u$ denotes the covariance matrix of $U$ which is assumed to be known. However, before minimizing $\sum_{i=1}^n(Y_i-\alpha-\beta^T Z_i(\lambda))^2$ and following the classcial SIMEX road map, we minimize the following conditional expectation
    \begin{equation}\label{eq2.2}
      E\left[\sum_{i=1}^n(Y_i-\alpha-\beta^T Z_i(\lambda))^2|(\bY,\bZ)\right],
    \end{equation}
  where $\bY=(Y_1,\ldots, Y_n)$ and $\bZ=(Z_1,\ldots, Z_n)$. Since $V_i$'s are i.i.d. from $N(0,\Sigma_u)$ and independent of other random variables in the model, so the expectation (\ref{eq2.2}) equals
    \begin{equation*}
        \sum_{i=1}^n(Y_i-\alpha-\beta^T Z_i)^2+n\lambda\bbt^T\Sigma_{U}\bbt.
    \end{equation*}
  The minimizer of the above expression is simply
    \begin{equation}\label{eq2.3}
      \hat\beta(\lambda)=(S_{ZZ}+\lambda\Sigma_{U})^{-1}S_{YZ},\quad \hat\alpha(\lambda)=\overline Y-\hat\beta^T(\lambda)\overline X,
    \end{equation}
  and by choosing $\lambda=-1$, we immediately have the commonly used bias-corrected estimators or the SIMEX estimators derived using the NON-IID pseudo-errors. Note that here not only do we not need the simulation step, but also the extrapolation step is unnecessary.

  Encouraged by this simple example, we dive into some more complex regression models to see if the above method can be applied to more general regression setups. In the following we use $m(x)$ to denote a regression function, $\theta_0$ the true value of $\theta$, and the regression error is independent of the covariate $X$. \vskip 0.1in

  \noindent {\bf Example 1:} Suppose $m(x)=\exp(x^T\theta)$. Then
       \begin{eqarray*}
        && E[(Y-m(Z(\lambda),\theta))^2|Y,Z]=E[(Y-\exp(Z^T\theta+\sqrt{\lambda}U^T\theta))^2|Y,Z]\\
        &=& Y^2-2Y\exp(Z^T\theta)E\exp(\sqrt{\lambda}U^T\theta)+\exp(2Z^T\theta)E\exp(2\sqrt{\lambda}U^T\theta)\\
        &=& Y^2-2Y\exp(Z^T\theta)\exp(\lambda\theta^T\Sigma_u\theta/2)+\exp(2Z^T\theta)
           \exp(2\lambda\theta^T\Sigma_u\theta).
       \end{eqarray*}\\ \vskip -0.5in \noindent
  Therefore, we can estimate $\theta$ by the minimizer of the empirical version of the above conditional expectation with $\lambda$ replaced with $-1$,
    $$
     L_n(\theta):=\frac{1}{n}\sum_{i=1}^n\left[ Y_i^2-2Y_i\exp(Z_i^T\theta-\theta^T\Sigma_u\theta/2)+\exp(2Z_i^T\theta-2\theta^T\Sigma_u\theta)\right].
    $$
  To see why minimizing the above target function leads to a consistent estimator, we take the derivative of $L_n(\theta)$ with respect to $\theta$ and set it to $0$. That is
    \begin{equation}\label{eq2.4}
      \frac{1}{n}\sum_{i=1}^n Y_i(Z_i-\Sigma_u\theta)\exp(Z_i^T\theta)-\frac{1}{n}\sum_{i=1}^n(Z_i-2\Sigma_u\theta)\exp(2Z_i^T\theta-3\theta^T\Sigma_u\theta/2)=0.
    \end{equation}
  Note that, as $n\to\infty$, almost surely,
   \begin{eqarray*}
     &&\frac{1}{n}\sum_{i=1}^n Y_iZ_i\exp(Z_i^T\theta)\to EYZ\exp(Z^T\theta)=\exp(\theta^T\Sigma_u\theta/2)E\left[(X+\Sigma_u\theta)\exp\left((\theta+\theta_0)^TX\right)\right],\\
     && \frac{1}{n}\sum_{i=1}^n Y_i\exp(Z_i^T\theta)\to \exp(\theta^T\Sigma_u\theta/2)E\exp\left((\theta+\theta_0)^TX\right),\\
     && \frac{1}{n}\sum_{i=1}^n\exp(2Z_i^T\theta)\to E\exp\left(2\theta^TX\right)\exp(2\theta^T\Sigma_u\theta),\\
     && \frac{1}{n}\sum_{i=1}^nZ_i\exp(2Z_i^T\theta)\to E\left[(X+2\Sigma_u\theta)\exp\left(2\theta^TX\right)\right]\exp(2\theta^T\Sigma_u\theta).
   \end{eqarray*}\\ \vskip -0.4in \noindent
  Therefore, as $n\to\infty$, almost surely,
    \begin{eqarray*}
      &&\dot L_n(\theta)\to\\
      && \exp(\theta^T\Sigma_u\theta/2)E(X+\Sigma_u\theta)\exp\left((\theta+\theta_0)^TX\right) - \exp(\theta^T\Sigma_u\theta/2)E\exp\left((\theta+\theta_0)^TX\right)\Sigma_u\theta\\
       && -\exp(\theta^T\Sigma_u\theta/2)E(X+2\Sigma_u\theta)\exp\left(2\theta^TX\right)+2\exp(\theta^T\Sigma_u\theta/2)E\exp\left(2\theta^TX\right)\Sigma_u\theta.
    \end{eqarray*}\\ \vskip -0.5in \noindent
 Denote the limit as $\dot L(\theta)$. Then $\dot L(\theta)=0$ if and only if
    \begin{eqarray*}
      EX\exp\left((\theta+\theta_0)^TX\right) -EX\exp\left(2\theta^TX\right)=0.
    \end{eqarray*}\\ \vskip -0.5in \noindent
 Obviously, $\theta_0$ is a solution.\vskip 0.1in

  \noindent {\bf Example 2:}  Suppose $m(x)=\sin(x^T\theta)$. Then
       \begin{eqarray*}
        && E[(Y-m(Z(\lambda),\theta))^2|Y,Z]=E[(Y-\sin(Z^T\theta+\sqrt{\lambda}U^T\theta))^2|Y,Z]\\
        &=& Y^2-2Y\sin(Z^T\theta)E\cos(\sqrt{\lambda}U^T\theta)-2Y\cos(Z^T\theta)E\sin(\sqrt{\lambda}U^T\theta)\\
        & &+\sin^2(Z^T\theta)E\cos^2(\sqrt{\lambda}U^T\theta)+\cos^2(Z^T\theta)E\sin^2(\sqrt{\lambda}U^T\theta)\\
        & &+2\sin(Z^T\theta)\cos(Z^T\theta)E\cos(\sqrt{\lambda}U^T\theta)\sin(\sqrt{\lambda}U^T\theta)\\
        &=&  Y^2-2Y\sin(Z^T\theta)E\cos(\sqrt{\lambda}U^T\theta)\\
        & &+\sin^2(Z^T\theta)E\cos^2(\sqrt{\lambda}U^T\theta)+\cos^2(Z^T\theta)E\sin^2(\sqrt{\lambda}U^T\theta).
       \end{eqarray*}\\ \vskip -0.5in \noindent
  Note that
     $$
       E\cos(\sqrt{\lambda}U^T\theta)=E\exp\left(\mathbf{i}\sqrt{\lambda}U^T\theta\right)
       =\exp\left(-\frac{1}{2}\lambda\theta^T\Sigma_u\theta\right),
     $$
  and from the well known trigonometric identities $\sin^2u=(1-\cos(2u))/2$ and $\cos^2u=(1+\cos(2u))/2$, we have
    $$
     E\cos^2(\sqrt{\lambda}U^T\theta)=\frac{1}{2}+\frac{1}{2}E\cos(2\sqrt{\lambda}U^T\theta)
      =\frac{1}{2}+\frac{1}{2}\exp\left(-2\lambda\theta^T\Sigma_u\theta\right),
    $$
     $$
     E\sin^2(\sqrt{\lambda}U^T\theta)=\frac{1}{2}-\frac{1}{2}E\cos(2\sqrt{\lambda}U^T\theta)
      =\frac{1}{2}-\frac{1}{2}\exp\left(-2\lambda\theta^T\Sigma_u\theta\right).
    $$
  Therefore, by $\cos^2u-\sin^2u=\cos(2u)$, we get
     \begin{eqarray*}
        &&E[(Y-m(Z(\lambda),\theta))^2|Y,Z]\\
        &=&Y^2-2Y\sin(Z^T\theta)\exp\left(-\frac{1}{2}\lambda\theta^T\Sigma_u\theta\right)
           -\frac{1}{2}\cos(2Z^T\theta)\exp\left(-2\lambda\theta^T\Sigma_u\theta\right)+1,
     \end{eqarray*}\\ \vskip -0.5in \noindent
  and an estimator of $\theta$ can be obtained by minimizing
    $$
     \frac{1}{n}\sum_{i=1}^n\left[ Y_i^2-2Y_i\sin(Z_i^T\theta)\exp\left(\frac{1}{2}\theta^T\Sigma_u\theta\right)
           -\frac{1}{2}\cos(2Z_i^T\theta)\exp\left(2\theta^T\Sigma_u\theta\right)\right]+1.
    $$
  By a simple algebra, we can show that as $n\to\infty$, the above average converges almost surely to $1/2+\sigma_\vep^2+E[\sin(X^T\theta_0)-\sin(X^T\theta)]^2$, which is minimized at $\theta=\theta_0$.
 \vskip 0.1in

  \noindent{\bf Example 3 (Poisson Regression):} Given a $p$-dimensional covariate $X$, suppose a nonnegative integer-valued random variable $Y$ has a Poisson distribution with mean $\exp(X^T\theta)$. If a sample $(X_i,Y_i), i=1,2,\ldots,n$ can be drawn from $(X,Y)$, then we can estimate $\theta$ by maximizing the log-likelihood function $\sum_{i=1}^n[Y_iX_i^T\theta-\exp(X_i^T\theta)-\log Y_i!]$. Note that
    \begin{eqarray*}
      E[YZ^T(\lambda)\theta-\exp(Z^T(\lambda)\theta)|(Y,Z)]
      &=&YZ^T\theta-\exp(Z^T\theta)E(\exp(\sqrt{\lambda}V^T\theta))\\
      &=&
      YZ^T\theta-\exp(Z^T\theta)\exp(\lambda\theta^T\Sigma_u\theta/2).
    \end{eqarray*}\\ \vskip -0.4in \noindent
  Thus, by directly extrapolating $\lambda$ to $-1$, we can estimate $\theta$ using the maximizer of
      $$\sum_{i=1}^n[Y_iZ_i^T\theta-\exp(Z_i^T\theta-\theta^T\Sigma_u\theta/2)]$$
  which coincides with the estimation procedure proposed in \cite{guo2002}.\vskip 0.1in

  The three examples discussed above demonstrate that in some cases, $\lambda=-1$ can be directly plugged into the conditional expectations and an estimator of $\theta$ can be obtained by solving a standard nonlinear equation system. However, this nice property is not shared by other models universally. \vskip 0.1in

  \noindent{\bf Example 4 (Logistic Regression):} In logistic regression, a $0$\,-\,$1$ random variable $Y$ depends on a $p$-dimensional covariate $X$ via the probability distribution $P(Y=1)=1-P(Y=0)=F(\alpha+X^T\beta)$, where $F(x)=1/(1+\exp(-x))$ is the logistic function. The log-likelihood function of $\alpha,\beta$ based on a sample $(Y_i, X_i)$, $i=1,2,\ldots, n$ is given by $\sum_{i=1}^n \left\{Y_i(\alpha+X_i^T\beta)-\log[1+\exp(\alpha+X_i^T\beta)]\right\}$.
  With $Z(\lambda)=Z+\sqrt{\lambda}V$, we have
    \begin{eqarray*}
      &&E[Y(\alpha+Z^T(\lambda)\beta)|(Y,Z)]-E\left\{\log[1+\exp(\alpha+Z^T(\lambda)\beta)]|(Y,Z)\right\}\\
      &=& Y(\alpha+Z^T\beta)-\int \log[1+\exp(\alpha+Z^T\beta+u)]\phi(u,0;\lambda\beta^T\Sigma_u\beta)du.
    \end{eqarray*}\\ \vskip -0.4in \noindent
  Note that $\lambda=-1$ can not be plugged into the integration.\vskip 0.1in

  \noindent {\bf Example 5:}  Consider the multiplicative regression model $Y=\exp(X^T\theta)\vep$, where $\vep$ is a positive random variable, with $E(\vep|X)=1$. This model is often called the accelerated failure time (AFT) model and is widely used in the survival analysis, econometrics and finance areas.
  There are two methods in the literature to estimate $\theta$ when observations can be made on $(Y,X)$. The first one is the least absolute relative error (LARE) estimation procedure which minimizes the following criterion
     $$
       LARE(\theta)=\sum_{i=1}^n\left[ Y_i^{-1}|Y_i-\exp(X_i^T\theta)|+\exp(-X_i^T\theta)|Y_i-\exp(X_i^T\theta)|\right]
     $$
  proposed by \cite{chen2010}. The second method is called the least product relative error (LPRE) estimation which minimizes the following expression
     $$
       LPRE(\theta)=\sum_{i=1}^n\left[Y_i\exp(-X_i^T\theta)+Y_i^{-1}\exp(X_i^T\theta)\right]
     $$
  proposed in \cite{chen2016}.

  In the case of $X$ being contaminated with normal measurement errors, according to the previous arguments, for the LARE criterion, we may consider the following conditional expectation:
    \begin{equation}\label{eq2.5}
     E\left\{\left[ Y^{-1}|Y-\exp(Z^T(\lambda)\theta)|+\exp(-Z^T(\lambda)\theta)|Y-\exp(Z^T(\lambda)\theta)|\right]\Big|(Y,Z)\right\}
    \end{equation}
  and for the LPRE criterion, we will calculate
    \begin{equation}\label{eq2.6}
      E\left\{\left[Y\exp(-Z^T(\lambda)\theta)+Y^{-1}\exp(Z^T(\lambda)\theta)\right]\Big|(Y,Z)\right\}.
    \end{equation}

  For (\ref{eq2.5}), we have
    \begin{eqarray*}
      && E\left\{Y^{-1}|Y-\exp(Z^T(\lambda)\theta)|\Big|(Y,Z)\right\}
      =\int\left\{Y^{-1}|Y-\exp(Z^T\theta+v)|\right\}\phi(v;0,\lambda\theta^T\Sigma_u\theta)dv\\
      &=& 2\Phi(\log(Y\exp(-Z^T\theta));0,\lambda\theta^T\Sigma_u\theta)-1\\
      && +Y^{-1}\exp(Z^T\theta+\lambda\theta^T\Sigma_u\theta/2)\left[1-2\Phi(\log(Y\exp(-Z^T\theta))+\lambda\theta^T\Sigma_u\theta,
      0, \lambda\theta^T\Sigma_u\theta)\right],
    \end{eqarray*}\\ \vskip -0.5in \noindent
  and
    \begin{eqarray*}
      && E\left\{\exp(-Z^T(\lambda)\theta)|Y-\exp(Z^T(\lambda)\theta)|\Big|(Y,Z)\right\}\\
      &=&\int\left\{\exp(-Z^T\theta+v)|Y-\exp(Z^T\theta+v)|\right\}\phi(v;0,\lambda\theta^T\Sigma_u\theta)dv\\
      &=& 1-2\Phi(\log(Y\exp(-Z^T\theta));0,\lambda\theta^T\Sigma_u\theta)\\
      && +Y\exp(-Z^T\theta+\lambda\theta^T\Sigma_u\theta/2)\left[2\Phi(\log(Y\exp(-Z^T\theta))+ \lambda\theta^T\Sigma_u\theta,0,\lambda\theta^T\Sigma_u\theta)-1\right].
    \end{eqarray*}\\ \vskip -0.5in \noindent
  Therefore, we have
   \begin{eqarray*}
     &&E\left\{\left[ Y^{-1}|Y-\exp(Z^T(\lambda)\theta)|+\exp(-Z^T(\lambda)\theta)|Y-\exp(Z^T(\lambda)\theta)|\right]\Big|(Y,Z)\right\}\nonumber\\
      &=& Y^{-1}\exp(Z^T\theta+\lambda\theta^T\Sigma_u\theta/2)\left[1-2\Phi(\log(Y\exp(-Z^T\theta))+\lambda\theta^T\Sigma_u\theta, 0, \lambda\theta^T\Sigma_u\theta)\right]\\
      && +Y\exp(-Z^T\theta+\lambda\theta^T\Sigma_u\theta/2)\left[2\Phi(\log(Y\exp(-Z^T\theta))+ \lambda\theta^T\Sigma_u\theta,0,\lambda\theta^T\Sigma_u\theta)-1\right].\nonumber
   \end{eqarray*}\\ \vskip -0.5in \noindent
  Note that we cannot directly plug $\lambda=-1$ into the above expectation, since $\lambda\theta^T\Sigma_u\theta$ serves as the variance of a normal distribution. However, for (\ref{eq2.6}), we have
    \begin{eqarray*}
     &&E\left[Y\exp(-Z^T(\lambda)\theta)+Y^{-1}\exp(Z^T(\lambda)\theta)\Big|(Y,Z)\right]\nonumber\\
      &=& \left[Y\exp(-Z^T\theta)+Y^{-1}\exp(Z^T\theta)\right]\exp\left(\lambda\theta^T\Sigma_u\theta/2\right).
   \end{eqarray*}\\ \vskip -0.5in \noindent
  $\lambda=-1$ can be directly plugged in. \vskip 0.1in

  \noindent {\bf Example 6 (Quantile Regression):} For a positive number $\tau\in (0,1)$, let $\rho_\tau(x)=x(\tau-I(x<0))$. Then the quantile regression estimates the regression coefficients $\beta$ by the minimizer of
  $\sum_{i=1}^n\rho_\tau(Y_i-X_i^T\beta)$ when both $Y$ and $X$ are observable. In the measurement error setup, similar to the previous examples, we may estimate $\beta$ by maximizing the conditional expectation
     $
        \sum_{i=1}^nE[\rho_\tau(Y_i-Z_i^T(\lambda)\beta)|(Y_i,Z_i)].
     $
  Calculation shows that
    \begin{eqarray*}
      && E[\rho_\tau(Y-Z^T(\lambda)\beta)|(Y,Z)]\\
      &=&E\left[(Y-Z^T\beta-\sqrt{\lambda}V^T\beta)(\tau-I(Y-Z^T\beta-\sqrt{\lambda}V^T\beta<0))|(Y,Z)\right]\\
      &=&\tau (Y-Z^T\beta)-(Y-Z^T\beta)\int_{Y-Z^T\beta}^\infty \phi(v;0,\lambda\beta^T\Sigma_u\beta)dv
         +\int_{Y-Z^T\beta}^\infty v\phi(v;0,\lambda\beta^T\Sigma_u\beta)dv\\
      &=&(\tau-1)(Y-Z^T\beta)+(Y-Z^T\beta)\Phi(Y-Z^T\beta;0,\lambda\beta^T\Sigma_u\beta)
      +\lambda\beta^T\Sigma_u\beta \phi(Y-Z^T\beta;0,\lambda\beta^T\Sigma_u\beta)].
    \end{eqarray*}\\ \vskip -0.4in \noindent
  Denote $\xi_i(\beta)=Y_i-Z_i^T\beta$. We can see that the target function has the form of
    $$
     (\tau-1)\sum_{i=1}^n\xi_i(\beta)+\sum_{i=1}^n\xi_i(\beta)\Phi(\xi_i(\beta);0,\lambda\beta^T\Sigma_u\beta)\\
      +\lambda\beta^T\Sigma_u\beta \sum_{i=1}^n\phi(\xi_i(\beta);0,\lambda\beta^T\Sigma_u\beta).
    $$
  It is easy to see that the new target function is a nonlinear differentiable function of $\beta$, and it can be readily minimized using standard algorithms.
  \vskip 0.1in

   \noindent {\bf Example 7 (Walsh Regression):} Another robust estimation procedure is the Walsh-average regression proposed in \cite{feng2012}. For a response variable $Y$ and a covariate vector $X$, the Walsh-average regression estimates the regression coefficient $\beta$ by minimizing the following objective function,
     $$
        \frac{1}{2n(n+1)}\sum_{i\leq j}|Y_i-X_i^T\beta+Y_j-X_j^T\beta|:=\frac{1}{2n(n+1)}\sum_{i\leq j} L(Y_i, Y_j, X_i, X_j;\beta)
     $$
  based on a sample $(Y_i, X_i), i=1,\ldots,n$ from $(Y, X)$. When $X_i$'s are measured with normal measurement error $U_i$, then similar to the previous arguments, we may estimate $\beta$ by maximizing the conditional expectation
     \begin{eqarray*}
        && \sum_{i\leq j}E[L(Y_i, Y_j, Z_i^T(\lambda),Z_i^T(\lambda);\beta)|(Y_i, Y_j, Z_i, Z_j)]\\
        &=& 2\sum_{i=1}^nE[|Y_i-Z_i^T(\lambda)\beta||(Y_i, Z_i)]+\sum_{i<j} E[|Y_i+Y_j-(Z_i+Z_j)^T(\lambda)\beta||(Y_i, Y_j, Z_i, Z_j)].
     \end{eqarray*}\\ \vskip -0.5in \noindent
  Based on the discussion on the quantile regression, taking $\tau=1/2$, the first term on the right-hand side equals
    \begin{eqarray*}
     2\sum_{i=1}^n\xi_i(\beta)[2\Phi(\xi_i(\beta),0,\lambda\beta^T\Sigma_u\beta)-1]
      +4\lambda\beta^T\Sigma_u\beta\sum_{i=1}^n\phi(\xi_i(\beta),0,\lambda\beta^T\Sigma_u\beta),
    \end{eqarray*}\\ \vskip -0.5in \noindent
  and the second term on the right-side equals
    \begin{eqarray*}
     \sum_{i<j}\xi_{ij}(\beta)[2\Phi(\xi_{ij}(\beta),0,2\lambda\beta^T\Sigma_u\beta)-1]
      +8\lambda\beta^T\Sigma_u\beta\sum_{i<j}\phi(\xi_{ij}(\beta),0,2\lambda\beta^T\Sigma_u\beta),
    \end{eqarray*}\\ \vskip -0.3in \noindent
  where $\xi_{ij}(\beta)=Y_i+Y_j-(Z_i+Z_j)^T\beta$.
  Again, this leads to a nonlinear target function to implement the proposed estimation procedure.
   \vskip 0.1in

   \noindent {\bf Example 8 (Expectile Regression):} For a positive number $\tau\in (0,1)$, let $\rho_\tau(x)=x^2[\tau-I(x<0)]$. Then the expectile regression estimates the parameter $\beta$ by minimizing the target function $\sum_{i=1}^n\rho_\tau(Y_i-X_i^T\beta)$ when both $Y$ and $X$ are observable. When $X$ is contaminated with normal measurement error, $\beta$ might be estimated by the maximizer of the conditional expectation
     $
        \sum_{i=1}^nE[\rho_\tau(Y_i-Z_i^T(\lambda)\beta)|(Y_i,Z_i)].
     $
  Calculation shows that
    \begin{eqarray*}
      && E[\rho_\tau(Y-Z^T(\lambda)\beta)|(Y,Z)]\\
      &=&E\left[(Y-Z^T\beta-\sqrt{\lambda}V^T\beta)^2[\tau-I(Y-Z^T\beta-\sqrt{\lambda}V^T\beta<0)]|(Y,Z)\right]
    \end{eqarray*}
    \begin{eqarray*}
      &=&\int (Y-Z^T\beta-v)^2|\tau-I(Y-Z^T\beta-v<0)|\phi(v;0,\lambda\beta^T\Sigma_u\beta)dv\\
      &=&\tau \int_{-\infty}^{Y-Z^T\beta}[(Y-Z^T\beta)^2-2v(Y-Z^T\beta)+v^2]\phi(v;0,\lambda\beta^T\Sigma_u\beta)dv\\
      & & +(1-\tau)\int_{Y-Z^T\beta}^\infty [(Y-Z^T\beta)^2-2v(Y-Z^T\beta)+v^2]\phi(v;0,\lambda\beta^T\Sigma_u\beta)dv\\
      &=&\tau (Y-Z^T\beta)^2\Phi(Y-Z^T\beta; 0,\lambda\beta^T\Sigma_u\beta)-2\tau (Y-Z^T\beta)\int_{-\infty}^{Y-Z^T\beta}v\phi(v;0,\lambda\beta^T\Sigma_u\beta)dv\\
      &&+\tau  \int_{-\infty}^{Y-Z^T\beta}v^2\phi(v;0,\lambda\beta^T\Sigma_u\beta)dv+(1-\tau) (Y-Z^T\beta)^2[1-\Phi(Y-Z^T\beta; 0,\lambda\beta^T\Sigma_u\beta)]\\
      & &-2(1-\tau) (Y-Z^T\beta)\int_{Y-Z^T\beta}^\infty v\phi(v;0,\lambda\beta^T\Sigma_u\beta)dv +(1-\tau)  \int_{Y-Z^T\beta}^{\infty} v^2\phi(v;0,\lambda\beta^T\Sigma_u\beta)dv\\
      &=&\tau (Y-Z^T\beta)^2\Phi(Y-Z^T\beta; 0,\lambda\beta^T\Sigma_u\beta)+2\tau \lambda\beta^T\Sigma_u\beta (Y-Z^T\beta)\phi(Y-Z^T\beta;0,\lambda\beta^T\Sigma_u\beta)\\
      &&-\tau \lambda\beta^T\Sigma_u\beta [(Y-Z^T\beta)\phi(v;0,\lambda\beta^T\Sigma_u\beta)-\Phi(v;0,\lambda\beta^T\Sigma_u\beta)] \\
      &&+(1-\tau) (Y-Z^T\beta)^2[1-\Phi(Y-Z^T\beta; 0,\lambda\beta^T\Sigma_u\beta)]\\
      & &-2(1-\tau)\lambda\beta^T\Sigma_u\beta (Y-Z^T\beta)\phi(Y-Z^T\beta;0,\lambda\beta^T\Sigma_u\beta)\\
      && +(1-\tau)\lambda\beta^T\Sigma_u\beta [(Y-Z^T\beta)\phi(Y-Z^T\beta;0,\lambda\beta^T\Sigma_u\beta)+1-\Phi(Y-Z^T\beta;0,\lambda\beta^T\Sigma_u\beta)]\\
      &=& (2\tau-1)(Y-Z^T\beta)^2\Phi(Y-Z^T\beta; 0,\lambda\beta^T\Sigma_u\beta)\\
      & & +(2\tau-1)\lambda\beta^T\Sigma_u\beta (Y-Z^T\beta)\phi(Y-Z^T\beta; 0,\lambda\beta^T\Sigma_u\beta)\\
      & & +(2\tau-1)\lambda\beta^T\Sigma_u\beta \Phi(Y-Z^T\beta; 0,\lambda\beta^T\Sigma_u\beta)  +(1-\tau)[(Y-Z^T\beta)^2+\lambda\beta^T\Sigma_u\beta].
    \end{eqarray*}\\ \vskip -0.5in \noindent
  Recall the notations $\xi_i(\beta)$, we can see that the target function takes the form of
    \begin{align*}
     &(2\tau-1)\sum_{i=1}^n\xi_i^2(\beta)\Phi(\xi_i(\beta); 0,\lambda\beta^T\Sigma_u\beta)+(2\tau-1)\lambda\beta^T\Sigma_u\beta\sum_{i=1}^n\xi_i(\beta)\phi(\xi_i(\beta); 0,\lambda\beta^T\Sigma_u\beta)\\
     &+(2\tau-1)\lambda\beta^T\Sigma_u\beta\sum_{i=1}^n\Phi(\xi_i(\beta); 0,\lambda\beta^T\Sigma_u\beta)+(1-\tau)\sum_{i=1}^n[\xi_i^2(\beta)+\lambda\beta^T\Sigma_u\beta].
    \end{align*}
  Like the quantile regression in Example 6, the new target function for the expectile regression is also a nonlinear differentiable function of $\beta$, and it can be readily minimized using standard algorithms.
  \vskip 0.1in

  Most estimation procedures are built upon optimizing specific target functions or searching for the solution of certain estimating equations. The above interesting findings may suggest that, after replacing the true predictors with the pseudo-data in the target functions or the estimating equations, one can simply optimize the conditional expectation or solve the new estimating equations. This allows researchers to circumvent the computationally-intensive simulation step of the classical SIMEX procedure. Also, if the process goes smoothly, the conditional expectation can be directly extrapolated to $\lambda=-1$ and an estimate can be obtained by solving a standard nonlinear equation, as shown in Examples 1, 2, 3, as well as the LPRE procedure in Example 4. In cases such as those described in Examples 4-8, where directly plugging $\lambda=-1$ into the resulting expectation is not feasible, one can proceed with the extrapolation step to obtain an estimate for $\theta$. In Section \ref{sec3}, we shall formulate the extrapolation estimation procedure for the general parametric regression models, and discuss its statistical properties.

 \section{Extrapolation Estimation in Parametric Regression}\label{sec3}

 For a general parametric regression model $Y=m(X;\theta)+\vep$ with $Z=X+U$, where $X\in\R^p$, $\vep$, $X$, $U$ are independent, $\theta\in\Theta\subset \R^q$, and $p, q$ are some positive integers, we may have different ways to estimate $\theta$ based on various assumptions on the model. In this section, the least squares estimation (LSE) procedure will be used as an exemplary method to construct the extrapolation estimation procedure. That is, the following conditional expectation
    \begin{equation*}
      E\left[\sum_{i=1}^n(Y_i-m(Z_i(\lambda);\theta))^2|(\bY,\bZ)\right]=
      \sum_{i=1}^n\int(Y_i-m(Z_i+u;\theta))^2\phi(u,0,\lambda\Sigma_u)du
    \end{equation*}
 with respect to $\theta$ will be minimized. To see intuitively why extrapolating $\lambda$ to $-1$ can result in a reasonable estimate of $\theta_0$, the true value of $\theta$, in this general setup as the sample size $n\to\infty$, we denote $\theta(\lambda)=\mbox{argmin}_\theta L(\theta; \lambda)$, where
 $L(\theta; \lambda)=E \int (Y-m(Z+u;\theta))^2\phi(u,0,\lambda\Sigma_u)du$, and
   \begin{equation}\label{eq3.1}
     L_n(\theta; \lambda)=\frac{1}{n}\sum_{i=1}^n\int(Y_i-m(Z_i+u;\theta))^2\phi(u,0,\lambda\Sigma_u)du.
   \end{equation}
 Under some regularity conditions, by the strong law of large numbers, $L_n(\theta;\lambda)\Longrightarrow L(\theta;\lambda)$ almost surely as $n\to\infty$. This, together with the fact
   \begin{equation}\label{eq3.2}
     L(\theta; \lambda)=E\int(Y-m(X+u;\theta))^2\phi(u,0,(\lambda+1)\Sigma_u)du\Longrightarrow E(Y-m(X;\theta))^2
   \end{equation}
  as $\lambda\to -1$ if we assume that $m(x;\theta)$ is continuous in $x$ for each $\theta\in\Theta$, implies that $\theta(\lambda)\to\theta_0$ if the equation $E(m(X;\theta_0)-m(X;\theta))^2=0$ has a unique solution. This heuristic argument leads to the following extrapolation algorithm for estimating $\theta$:\vskip 0.2in
   \begin{center}
   \underline{~~~~~~~~~~~~~~~~~~~~~~~~~~~~~~~~~~~~~~~~~~~~~~~~~~~~~~~~~~~~~~~~~~~~~~~~~~~~~~~~~~~~~~~~~~~~~~~~~~~~~~~~~~~~~~~~~~}
   \vskip 0.1in
   \begin{minipage}[c]{5.5in}
      \hskip -0.3in {\bf Extrapolation Estimation Algorithm}\vskip 0.1in
       \begin{enumerate}
        \item If $L_n(\theta; \lambda)$ can be directly extrapolated to $\lambda=-1$, then an estimate of $\theta_0$ is given by $\hat\theta_n=\mbox{argmin}_\theta L_n(\theta; -1)$, where $L_n(\theta; \lambda)$ is defined in (\ref{eq3.1}).\vskip 0.05in

        \item If $L_n(\theta; \lambda)$ cannot be directly extrapolated to $\lambda=-1$, then pre-select some grid points $0=\lambda_1<\lambda_2<\cdots<\lambda_K=c$, for example, $c=2$.
      \begin{itemize}
        \item For each $\lambda\in\{\lambda_1,\ldots,\lambda_K\}$, solving
                $ \hat\theta_n(\lambda)=\mbox{argmin}_\theta L_n(\theta; \lambda)$.
        \item Fit a trend for the pairs $(\lambda_k, \hat\theta_n(\lambda_k))$, $k=1,2,\ldots,K$, and extrapolate this trend back to $-1$.
      \end{itemize}
    Take the extrapolated value $\hat\theta_n$ as an estimate of $\theta_0$.
     \end{enumerate}
   \end{minipage}\vskip 0.1in
   \underline{~~~~~~~~~~~~~~~~~~~~~~~~~~~~~~~~~~~~~~~~~~~~~~~~~~~~~~~~~~~~~~~~~~~~~~~~~~~~~~~~~~~~~~~~~~~~~~~~~~~~~~~~~~~~~~~~~~}
   \end{center}

  In the following we shall explore the extrapolation estimation algorithm described above in detail by presenting some theoretical results. For a generic parametric function $g(x;\theta)$ with multidimensional $x$ and $\theta$, we denote $\dot g(x;\theta)=\partial g(x; \theta)/\partial\theta$, $\ddot g(x;\theta)=\partial^2 g(x;\theta)/\partial\theta\partial\theta^T$, and $g'(x;\theta)=\partial g(x; \theta)/\partial x$, $ g''(x;\theta)=\partial^2 g(x;\theta)/\partial x\partial x^T$.\vskip 0.1in

  First we list some technical assumptions needed for presenting the theoretical arguments.
    \begin{itemize}
       \item[] {\bf (C1)}. The parameter space $\Theta$ of $\theta$ is compact;
       \item[] {\bf (C2)}. For each $x\in\R^p$, the regression function $m(x; \theta)$ is twice continuously differentiable for each $\theta\in\Theta$;
       \item[] {\bf (C3)}. There exists a function $K(x)$, not depending on $\theta$, such that $EK^2(Z+V)<\infty$ and
            $ \left|\ddot{m}(x;\theta)\right|+|m(x;\theta)|\leq K(x)$ for all $x$ in the domain of $m$;
       \item[] {\bf (C4)}. For each $\lambda\geq 0$, the minimizer of $L(\theta,\lambda)$ exists and is unique;
       \item[] {\bf (C5)}. $\argmin_{\theta\in\Theta}E[m(X;\theta)-m(X;\theta_0)]^2$ is unique;
       \item[] {\bf (C6)}. For each $\lambda\geq 0$,
           $E\left[\dot m(Z+V;\theta)\dot m^T(Z+V;\theta)\right]$
            is positive definite, where $V\sim N(0,\lambda\Sigma_u)$, $Z$ and $V$ are independent.
    \end{itemize}

  Conditions(C1)-(C3) allow us to show the uniform convergence of $L_n(\theta)$ to $L(\theta)$ over $\theta\in\Theta$ using the uniform convergence theory discussed in \cite{ferg1996}, which, together with the condition (C4), implies the convergence of $\hat\theta_n(\lambda)$ to $\theta(\lambda)$ as $n\to\infty$ for each $\lambda>0$, and eventually to $\theta_0$ by letting $\lambda\to -1$ by using (C5). In fact, if $\mbox{\rm argmin}_{\theta} L(\theta; \lambda)$ is not unique, then we can show that for any local minimizer, there is a sequence of minimizers of $L_n(\theta; \lambda)$ that converges to the local minimizer in probability. However, to keep the argument relatively simple, we shall adopt (C5) in the following discussion. Finally, using (C6), we can show the asymptotic normality of the extrapolation estimator. The following theorem summarizes the consistency of $\hat\theta_n(\lambda)$ to $\theta(\lambda)$ for each $\lambda>0$, and the approximation of $\theta(\lambda)$ to $\theta_0$ as $\lambda$ is extrapolated to $-1$.

  \begin{thm}\label{thm1} Suppose that the conditions (C1), (C2), (C3), and (C4) hold. Then for each $\lambda>0$, $\hat\theta_n(\lambda)\to\theta(\lambda)$ in probability as $n\to\infty$. If we further assume that (C5) holds, then as $\lambda\to -1$,
    \begin{eqarray*}
      \theta(\lambda)&=&\theta_0-(\lambda+1)\left[E\dot m(X;\theta_0)\dot m^T(X;\theta_0)\right]^{-1}\cdot\\
      && \hskip 0.2in\left[E\dot m(X;\theta_0) \mbox{trace}(m''(X;\theta_0)\Sigma_u^{2})+E\dot m'(X;\theta_0)\Sigma_u^2 m'(X;\theta_0)
      \right]+o((\lambda+1)).
   \end{eqarray*}
  \end{thm}

  Denote $\Lambda=(\lambda_1, \ldots, \lambda_K)^T$, $\hat\theta_n(\Lambda)=(\hat\theta_n(\lambda_1), \ldots, \hat\theta_n(\lambda_K))$, $\theta(\Lambda)=(\theta(\lambda_1), \ldots, \theta(\lambda_K))$. The asymptotic joint normality of $\hat\theta_n(\Lambda)$ to $\theta(\Lambda)$ is described in the following theorem.

  \begin{thm}\label{thm2} In addition to the conditions in Theorem \ref{thm1}, suppose (C6) holds. Then we have
   $$
    \sqrt{n}[(\hat\theta_n(\lambda_1)-\theta(\lambda_1))^T, \ldots, (\hat\theta_n(\lambda_K)-\theta(\lambda_K))^T]^T
    \sim N(0, \Omega_1^{-1}(\Lambda)\Omega_0(\Lambda)\Omega_1^{-1}(\Lambda)),
   $$
  where
   \begin{eqarray*}
     \Omega_0(\Lambda)&=&[\Sigma_0(\lambda_j,\lambda_k)]_{K\times K},\quad
     \Omega_1(\Lambda)=\mbox{Diag}(\Sigma_1(\lambda_1),\ldots, \Sigma_1(\lambda_K)),\\
     \Sigma_1(\lambda)&=&E\int \dot m(X+u;\theta(\lambda))\dot m^T(X+u;\theta(\lambda))\phi(u,0,(\lambda+1)\Sigma_u)du\\
       &&-E\int [m(X;\theta_0)-m(X+u,\theta(\lambda))]\ddot m(X+u;\theta(\lambda))\phi(u,0,(\lambda+1)\Sigma_u)du
    \end{eqarray*}\\ \vskip -0.5in \noindent
  and
  $\Sigma_0(\lambda_j,\lambda_l)$, $j, l=1,2,\ldots,K$ is defined as
   \begin{eqarray*}
      &&E\iint \left\{\sigma_\vep^2+[m(X,\theta_0)-m(X+u,\theta(\lambda_j))][m(X,\theta_0)-m(X+v,\theta(\lambda_l))]\right\}\dot m(X+u;\theta(\lambda_j)) \\
      &&\hskip 0.5in \dot m^T(X+v;\theta(\lambda_l))\phi(u,v,(\lambda_j+\lambda_l)\Sigma_u)\phi\left(\frac{\lambda_jv+\lambda_lu}{\lambda_j+\lambda_l},0,
     \frac{\lambda_{j}+\lambda_l+\lambda_j\lambda_l}{\lambda_j+\lambda_l}\Sigma_u\right)dudv
    \end{eqarray*}\\ \vskip -0.5in \noindent
 or
   \begin{eqarray*}
      &&E\iint \left\{\sigma_\vep^2+[m(X,\theta_0)-m(X+u,\theta(\lambda_j))][m(X,\theta_0)-m(X+v,\theta(\lambda_l))]\right\}\dot m(X+u;\theta(\lambda_j)) \\
      &&\hskip 0.5in \dot m^T(X+v;\theta(\lambda_l))\phi(v,0,(\lambda_l+1)\Sigma_u)\phi\left(u,\frac{v}{\lambda_l+1},\frac{\lambda_{1}+\lambda_l+\lambda_j\lambda_l}{\lambda_l+1}
     \Sigma_u\right)dudv.
   \end{eqarray*}
  \end{thm}

From Theorem \ref{thm2}, one can see that for each $\lambda>0$,
         \begin{equation*}
           \sqrt{n}(\hat\theta_n(\lambda)-\theta(\lambda))\to N\left(0,\Sigma_1^{-1}(\lambda)\Sigma_0(\lambda,\lambda)\Sigma_1^{-1}(\lambda)\right)
         \end{equation*}
in distribution.
%
%
%
 In particular, for $\lambda=0$, we have
  $$ \sqrt{n}(\hat\theta_n(0)-\theta(0))\to N(0,\Sigma_1^{-1}(0)\Sigma_0(0,0)\Sigma_1^{-1}(0)),$$
 where $\theta(0)=\argmin_{\theta\in\Theta}E[m(X;\theta_0)-m(Z;\theta)]^2$, and
   \begin{eqarray*}
      \Sigma_0(0,0)&=&E \left\{\sigma_\vep^2+[m(X,\theta_0)-m(Z,\theta(0))][m(X,\theta_0)-m(Z,\theta(0))]\right\}\dot m(Z;\theta(0))\dot m^T(Z;\theta(0)),\\
      \Sigma_1(0)&=&E\dot m(Z;\theta(0))\dot m^T(Z;\theta(0))-E[m(X;\theta_0)-m(Z,\theta(0))]\ddot m(Z;\theta(0)).
   \end{eqarray*}\\ \vskip -0.5in \noindent
 This recovers the asymptotic normality result for the naive estimator $\hat\theta_n(0)$ of $\theta_0$. It is in our interest to know the limits of $\Sigma_1(\lambda)$ and $\Sigma_0(\lambda,\lambda)$ when $\lambda\to -1$. The trend of $\Sigma_1(\lambda)$ is easy to derive. In fact, from the expression $\Sigma_1(\lambda)$ in Theorem \ref{thm2}, we have $\Sigma_1(-1)=E\dot m(X; \theta_0)\dot m^T(X;\theta_0)$. Let $\lambda_j=\lambda_l=\lambda$ in the second expression of $\Sigma_0(\lambda_j,\lambda_l)$ in Theorem \ref{thm2}, we can rewrite $\Sigma_0(\lambda,\lambda)$ as
    \begin{eqarray*}
      &&E\iint \left\{\sigma_\vep^2+[m(X,\theta_0)-m(X+u,\theta(\lambda))][m(X,\theta_0)-m(X+v,\theta(\lambda))]\right\}\dot m(X+u;\theta(\lambda)) \\
      &&\hskip 0.5in \dot m^T(X+v;\theta(\lambda))\phi(v,0,(\lambda+1)\Sigma_u)\phi\left(u,\frac{v}{\lambda+1},\frac{2\lambda+\lambda^2}{\lambda+1}
     \Sigma_u\right)dudv.
   \end{eqarray*}\\ \vskip -0.5in \noindent
 However, $\Sigma_0(\lambda,\lambda)$ does not have an explicit limit when $\lambda\to -1$, unless some strong conditions are imposed on the regression function.

 Following are two examples of $\Sigma_0(\lambda, \lambda)$ to further illustrate the above findings.\vskip 0.1in

 \noindent{\bf Example 1.} Suppose $m(x,\theta)=x'\theta$, then we have $\dot m(x,\theta)=x$. From the second expression of $\Sigma_0(\lambda,\lambda)$, we have
    \begin{eqarray*}
      \Sigma_0(\lambda,\lambda)&=&\sigma_\vep^2 E\iint(X+u)(X+v)^T\phi(v,0,(\lambda+1)\Sigma_u)
      \phi\left(u,\frac{v}{\lambda+1},\frac{\lambda(\lambda+2)}{\lambda+1}\Sigma_u\right)dudv\\
      & &+E\iint [X^T\theta_0-(X+u)^T\theta(\lambda)][X^T\theta_0-(X+v)^T\theta(\lambda)]\cdot\\
      & & \hskip 0.5in (X+u)(X+v)^T\phi(v,0,(\lambda+1)\Sigma_u)
      \phi\left(u,\frac{v}{\lambda+1},\frac{\lambda(\lambda+2)}{\lambda+1}\Sigma_u\right)dudv\\
      &=&  \sigma_\vep^2(EXX^T+\Sigma_u)+E(\theta_0-\theta(\lambda))^TXX^T(\theta_0-\theta(\lambda))XX^T\\
      &&   + (\theta_0-\theta(\lambda))^TE(XX^T)(\theta_0-\theta(\lambda))\Sigma_u + \theta(\lambda)^T\Sigma_u\theta(\lambda)E(XX^T)\\
      & & -E(\theta_0-\theta(\lambda))^TXX\theta(\lambda)^T \Sigma_u-(\lambda+1)E(\theta_0-\theta(\lambda))^TX \Sigma_u\theta(\lambda)X^T\\
      & & +\int v^T\theta(\lambda)\left[vv^T+\lambda(\lambda+2)\Sigma_u\right]\theta(\lambda)v^T
      \phi(v,0,\Sigma_u)dv.
   \end{eqarray*}\\ \vskip -0.5in \noindent
 Extrapolating $\lambda$ to $-1$, we see that $\Sigma_0(\lambda,\lambda)$ converges to
   \begin{equation}\label{eq3.3}
     \sigma_\vep^2(EXX^T+\Sigma_u)+\theta_0^T\Sigma_u\theta_0EXX^T+\int v^T\theta_0vv^T\theta_0v^T
      \phi(v,0,\Sigma_u)dv-\Sigma_u\theta_0\theta_0^T\Sigma_u.
   \end{equation}
  (\ref{eq3.3}) is exactly same as the asymptotic covariance matrix of the bias corrected estimator $\hat\beta(\lambda)$ defined in (\ref{eq2.3}) after taking $\lambda=-1$. Please refer to the Theorem 3.1 in \cite{liang1999} for a proof of the asymptotic normality of the bias corrected estimator, as well as its asymptotic covariance matrix (\ref{eq3.3}). \vskip 0.1in

 \noindent {\bf Example 2:} Consider the exponential regression function  $Y=\exp(\theta X)+\vep$ with a univariate predictor $X$. Note that $\dot m(x; \theta)=x \exp(\theta x)$.  It is easy to see that $\Sigma_1(\lambda)\to EX^2\exp(2X\theta_0)=\tau_1$ as $\lambda\to -1$.
  From the second expression for $\Sigma_0(\lambda, \lambda)$ in Theorem \ref{thm2}, we can rewrite it as the sum of the following two terms:
    \begin{eqarray*}
      S_1(\lambda)&=&\sigma_\vep^2E\iint(X+u)(X+v) \exp{[(X+u)\theta(\lambda)]}\exp{[\theta(\lambda)(X+v)]}\cdot \\
      & & \hskip 0.5in \phi(v,0,(\lambda+1)\sigma_u^2)
      \phi\left(u,\frac{v}{\lambda+1},\frac{\lambda(\lambda+2)}{\lambda+1}\sigma_u^2\right)dudv,\\
      S_2(\lambda)&=&
      E\iint \{\exp{(\theta_0 X)} - \exp{[\theta(\lambda)(X+u)]}\}\cdot
       \{\exp{(\theta_0 X)} - \exp{[\theta(\lambda)(X+v)]}\}\cdot \\
      & & \hskip 0.5in (X+u)(X+v) \exp{[(X+u)\theta(\lambda)]}\exp{[\theta(\lambda)(X+v)]}\cdot \\
      & & \hskip 0.5in \phi(v,0,(\lambda+1)\sigma_u^2)
      \phi\left(u,\frac{v}{\lambda+1},\frac{\lambda(\lambda+2)}{\lambda+1}\sigma_u^2\right)dudv.
   \end{eqarray*}\\ \vskip -0.5in \noindent
  A tedious computation shows that
    $$
      S_1(\lambda)= \sigma_\vep^2E\left\{(\sigma_u^2+[X+(\lambda+2)\theta(\lambda)\sigma_u^2]^2)\exp[2X\theta(\lambda)
      +(\lambda+2)\theta^2(\lambda)\sigma_u^2]\right\},
    $$
  and
    \begin{eqarray*}
      S_2(\lambda)&=&E X^2\exp[{2X(\theta_0+\theta(\lambda))+(\lambda+2)\sigma_u^2 \theta^2(\lambda)}] \\
      &&   - 2E X^2\exp[{X(\theta_0+3\theta(\lambda))+\frac{1}{2}(5\lambda+9)\sigma_u^2 \theta^2(\lambda)}]\\
      &&  + E X^2\exp[{4X\theta(\lambda)+4(\lambda+2)\sigma_u^2 \theta^2(\lambda)}]\\
       &&  + 2(\lambda+2)\sigma_u^2\theta(\lambda) E X\exp[{2X(\theta_0+\theta(\lambda))+(\lambda+2)\sigma_u^2 \theta^2(\lambda)}] \\
       &&   - 6(\lambda+2)\sigma_u^2\theta(\lambda)E X\exp[{X(\theta_0+3\theta(\lambda))+\frac{1}{2}(5\lambda+9)\sigma_u^2 \theta^2(\lambda)}]\\
      &&  + 4(\lambda+2)\sigma_u^2\theta(\lambda) E X\exp[{4X\theta(\lambda)+4(\lambda+2)\sigma_u^2 \theta^2(\lambda)}]\\
      &&  + [\sigma_u^2+(\lambda+2)^2\sigma_u^4\theta^2(\lambda)] E \exp[{2X(\theta_0+\theta(\lambda))+(\lambda+2)\sigma_u^2 \theta^2(\lambda)}] \\
       &&   - 2[\sigma_u^2+(2\lambda+3)(\lambda+3)\sigma_u^4\theta^2(\lambda)]E \exp[{X(\theta_0+3\theta(\lambda))+\frac{1}{2}(5\lambda+9)\sigma_u^2 \theta^2(\lambda)}]\\
      &&  + [\sigma_u^2+4(\lambda+2)^2\sigma_u^4\theta^2(\lambda)] E \exp[{4X\theta(\lambda)+4(\lambda+2)\sigma_u^2 \theta^2(\lambda)}].
    \end{eqarray*}\\ \vskip -0.5in \noindent
 Extrapolating $\lambda$ to $-1$, we see that $\Sigma_0(\lambda,\lambda)$ tends to
   \begin{eqarray*}
     \tau_0^2&=&\sigma_\vep^2E\left\{(\sigma_u^2+[X+\theta_0\sigma_u^2]^2)\exp[2X\theta_0
      +\theta_0^2\sigma_u^2]\right\}\\
      &&+E\left\{(\sigma_u^2+[X+\theta_0\sigma_u^2]^2)\exp[4X\theta_0
      +\theta_0^2\sigma_u^2]\right\}\\
      &&+E\left\{[\sigma_u^2+(X+2\theta_0\sigma_u^2)^2]\exp(4X\theta_0+4\theta_0^2\sigma_u^2)\right\}\\
     &&  - 2E\left\{(\sigma_u^2+X^2+3X \theta_0\sigma_u^2 + 2\theta_0^2 \sigma_u^4)\exp(4X\theta_0 + 2\sigma_u^2\theta_0^2)\right\}.
    \end{eqarray*}\\ \vskip -0.5in \noindent
 In fact, we can verify that the asymptotic variance of the estimator defined by the solution of (\ref{eq2.4}) is exactly $\tau_0^2/\tau_1^2$.\vskip 0.1in

 It is noted that the exact extrapolation function $\theta(\lambda)$ is implicitly defined by the equation
 $\dot L(\theta; \lambda)= 0$, where $L(\theta; \lambda)$ is defined by (\ref{eq3.2}). In some special cases, such as the linear, the exponential and the Poisson regressions discussed in Examples 1, 2 and 3, the exact extrapolation function can be obtained by solving the above equation. However, the solution generally has no closed-form.

For simplicity, we assume that $\theta$ is one-dimensional. By a Taylor expansion of $m(x+\sqrt{\lambda+1}\sigma_uu, \theta)$ and $\dot m(x+\sqrt{\lambda+1}\sigma_uu, \theta)$ at $\lambda=-1$ and $\theta$ at $\theta=\theta_0$, and plugging these two Taylor expansions in the expression of $\dot L(\theta; \lambda)$, and after some algebra, the solution of the equation $\dot L(\theta; \lambda)= 0$ has the form of
   $$
     \theta(\lambda)=\theta_0+\frac{a_0 (\lambda+1)}{a_1+a_2(\lambda+1)+a_3(\lambda+1)^2+\cdots},
   $$
  where $a_0, a_1, a_2, a_3, \ldots$ are some model-dependent constants. The above exact extrapolation function can be simplified to obtain several approximate simple functions to implement the extrapolation. For example, by truncating the denominator to the first order of $\lambda+1$, and after some equivalent transformation, we obtain the commonly used nonlinear extrapolation function $a+b/(c+\lambda)$; or if we apply another Taylor expansion for the denominator at $\lambda=-1$ up to first order, second order etc., then we can also obtain the linear, quadratic extrapolation function and so on.

 Almost all literature involving SIMEX assumes that the true extrapolation function has a known parametric form when discussing the asymptotic distributions of the SIMEX estimators. However, the true extrapolation function is generally unknown except for some special cases, and this discouraging observation really nullifies all the relevant theoretical developments based on known extrapolation functions. Unfortunately this is also true for our proposed extrapolation method.

 If we are fortunate to have a closed-form extrapolation function with a parametric form $G(\lambda,\Gamma)$, which is twice continuously differentiable with respect to the parameter $\Gamma\in\R^d$ for some positive integer $d$, then assuming that the true value of the parameter is $\Gamma_0$, that is $\theta_0=G(-1,\Gamma_0)$, we can estimate $\Gamma_0$ by minimizing the least squares criterion
   \begin{equation}\label{eq3.4}
     \|\hat\theta_n(\Lambda)-G(\Lambda,\Gamma)\|_2^2=\sum_{j=1}^K\|G(\lambda_j, \Gamma)-\hat\theta_n(\lambda_j)\|_2^2,
   \end{equation}
 where
  $
    G(\Lambda, \Gamma)=[G^T(\lambda_1,\Gamma), G^T(\lambda_2,\Gamma),\ldots,G^T(\lambda_K,\Gamma)]^T_{qK\times 1},
  $
 or solving the equation $$\dot G^T(\Lambda,\Gamma)(\hat\theta_n(\Lambda)-G(\Lambda,\Gamma))=0,$$ where
  $$
    \dot G(\Lambda,\Gamma)=[\dot G^T(\lambda_1,\Gamma), \dot G^T(\lambda_2,\Gamma),\ldots,\dot G^T(\lambda_K,\Gamma)]^T_{qK\times d},
  $$
 and for $j=1,2,\ldots, K$,
  $$
    \dot G(\lambda_j,\Gamma)=\left(\partial G(\lambda_j,\Gamma)/\partial\gamma_l\right)_{q\times d}.
  $$
 Denote the minimizer of (\ref{eq3.4}) as $\hat\Gamma$, then the resulting extrapolation estimator of $\theta_0$ will be $\hat{\theta}_n=G(-1,\hat\Gamma)$.
 Denote $H(\Lambda)=\dot G^T(\Lambda,\Gamma_0)\dot G(\Lambda,\Gamma_0)$ and
  \begin{eqarray*}
   \Pi(\Lambda)&=&H^{-1}(\Lambda)\dot G^T(\Lambda,\Gamma_0)\Omega_1^{-1}(\Lambda)\Omega_0(\Lambda)\Omega_1^{-1}(\Lambda)\dot G(\Lambda,\Gamma_0)H^{-1}(\Lambda),
  \end{eqarray*}\\ \vskip -0.5in \noindent
 We have the following ideal result.

\begin{thm}\label{thm3} Assuming that conditions (C1)-C(6) hold, and the true extrapolation function has a known parametric form $G(\lambda,\Gamma)$ with nonsingular $H(\Lambda)$, then
  \begin{eqarray*}
  \sqrt{n}(\hat{\theta}_n-\theta_{0})\Longrightarrow N(0,\dot G(-1,\Gamma_0)\Pi(\Lambda)\dot G^T(-1,\Gamma_0)).
  \end{eqarray*}
\end{thm}

\section{Numerical Studies}\label{sec4}

  In this section, we conduct some simulation studies for two parametric models to evaluate the finite sample performance of the proposed SIMEX procedure. In the first simulation study, we generated the data from a regression model with exponential regression function as described in Example 1. In the second simulation study, the data are generated from a bivariate quantile regression model as described in Example 6. Note that we can directly plug $\lambda=-1$ in the target function from the exponential regression model, but we cannot do the same in the quantile regression model, where an approximation is needed by using extrapolation functions. We also analyze a dataset from the National Health and Nutrition Examination Survey (NHANES) to illustrate the application of the proposed estimation procedure.  \vskip 0.1in

 \subsection{Simulation Studies}

    \noindent{\it Exponential Regression.} Consider a univariate exponential regression model first. Suppose $X$ is a one-dimensional standard normal random variable. From Section 2, we know the true value $\theta_0$ is a solution of the equation $EX\exp((\theta+\theta_0)X)-EX\exp(2\theta X)=0$. Calculation shows that the equation can be rewritten as
       $$
        (\theta+\theta_0)\exp((\theta+\theta_0)^2/2)-2\theta\exp(2\theta^2)=0.
       $$
    It is easy to see that $\theta_0$ is a solution of the above equation. Moreover, the solution of the above equation is also unique. Therefore, when the sample size is sufficiently large, the minimizer of $L_n(\theta)$ is unique. We generate $Z$ from $Z=X+U$ with $U\sim N(0, \sigma_u^2)$. To see the effect of the measurement error variance on the estimation procedure, we choose $\sigma^2=0.1, 0.25, 0.5$. The true parameter values are chosen to be $\theta=1$, and the sample sizes are chosen to be $n=200, 300, 500$ and $800$. For each setup, we repeat the simulation 500 times, and the means, biases, variances and the mean squared errors (MSE) are calculated and used to evaluate the finite sample performance of the proposed SIMEX procedure.
    \begin{table}{\footnotesize
      \begin{tabular}{c|ccc|ccc|ccc|ccc}
      \hline\hline
       & \multicolumn{3}{c}{n=200} \vline& \multicolumn{3}{c}{n=300} \vline& \multicolumn{3}{c}{n=500} \vline& \multicolumn{3}{c}{n=800}\\ \hline
 $\sigma_u^2$ &   0.5 &  0.25 & 0.1 & 0.5 & 0.25 & 0.1 & 0.5 & 0.25 & 0.1 & 0.5 & 0.25 &  0.1 \\ \hline
  Mean     & 1.085 & 1.057 & 1.025 & 1.067 & 1.054 & 1.017 & 1.037 & 1.033 & 1.016 & 1.041 & 1.024 & 1.013\\
  Bias     & 0.085 & 0.057 & 0.025 & 0.067 & 0.054 & 0.017 & 0.037 & 0.033 & 0.016 & 0.041 & 0.024 & 0.013\\
  Variance & 0.022 & 0.014 & 0.004 & 0.015 & 0.009 & 0.003 & 0.009 & 0.006 & 0.002 & 0.006 & 0.004 & 0.002\\
  MSE      & 0.029 & 0.017 & 0.005 & 0.019 & 0.012 & 0.003 & 0.010 & 0.007 & 0.003 & 0.008 & 0.005 & 0.002\\
  \hline\hline
      \end{tabular}}
      \caption{Mean, bias, variance and MSE of $\hat\theta$ for univariate exponential regression}\label{tab1}
    \end{table}

\begin{figure}[h!]
      \includegraphics[width=5in,height=4in]{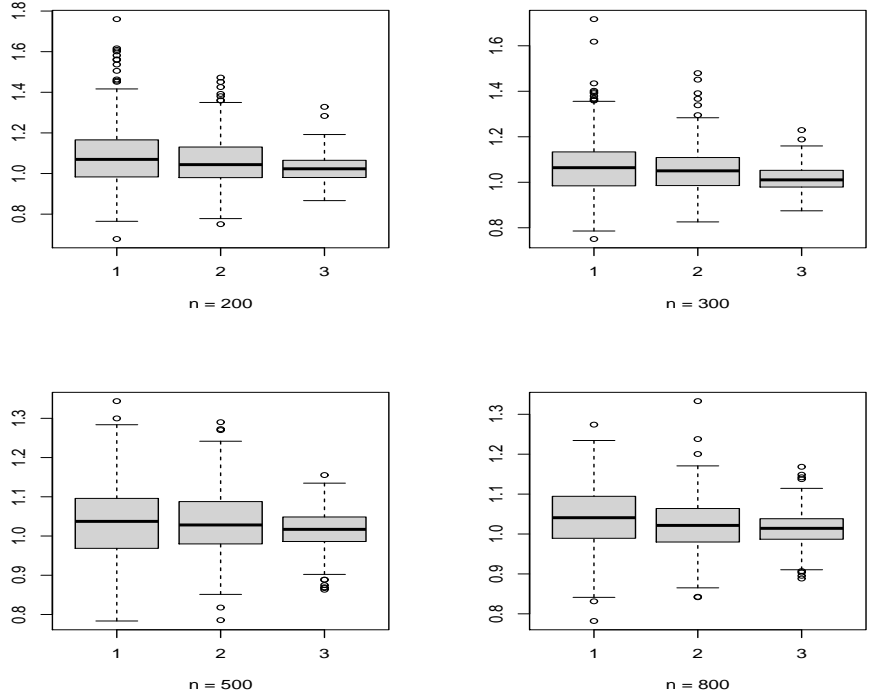}
      \caption{Boxplots of estimates $\hat\theta$ for univariate exponential regression}\label{fig1}
    \end{figure}

   Table \ref{tab1} and Figure \ref{fig1} show that when the sample size gets bigger, the performance of the proposed estimator becomes better, as evidenced by the decreasing biases, variances and MSEs. For a fixed sample size, it is seen that the smaller the measurement error variance, the smaller the bias, variance and MSE, as expected.

    We also conducted a simulation study for an exponential regression with two predictors $X=(X_1, X_2)$, that is, $m(x; \theta)=\exp(x^T\theta)$, $Z_1=X_1+U_1, Z_2=X_2+U_2$, where $\theta=(\beta_1, \beta_2)$. We choose $\vep$ to be standard normal, $(X_1, X_2)$ has a bivariate standard normal distribution, the measurement error $(U_1, U_2)$ is generated from a bivariate normal distribution with mean vector of $0$'s, equal variances $\sigma^2$ and covariance $0.5\sigma^2$. To see the effect of the measurement variance on the estimation procedure, we choose $\sigma^2=0.25, 0.2, 0.1$. The true parameter values are chosen to be $\beta_1=0.5$, $\beta_2=1$, and the sample sizes are chosen to be $n=200, 300, 500$ and $800$. Similar to the one dimensional case, for each setup, we repeat the simulation 500 times, and the means, biases, variances and the MSE are calculated and used to evaluate the finite sample performance of the proposed SIMEX procedure. Findings similar to the univariate case are observed.

    \begin{table}[h!]{\footnotesize
      \begin{tabular}{c|cccccc|cccccc}
      \hline\hline
      & \multicolumn{6}{c}{n=200} \vline& \multicolumn{6}{c}{n=300} \\ \hline
      $\sigma_u^2$ &  \multicolumn{2}{c}{0.25}  &
      \multicolumn{2}{c}{0.2}   &
      \multicolumn{2}{c}{0.1}  \vline &
      \multicolumn{2}{c}{0.25}  &
      \multicolumn{2}{c}{0.2}  &
      \multicolumn{2}{c}{0.1}  \\ \hline
      Mean     & 0.690 & 1.295 & 0.623 & 1.164 & 0.540 & 1.050 & 0.659 & 1.195 & 0.615 & 1.101 & 0.534 &1.044\\
      Bias     & 0.190 & 0.295 & 0.123 & 0.164 & 0.040 & 0.050 & 0.159 & 0.195 & 0.115 & 0.101 & 0.034 &0.044\\
      Variance & 0.079 & 0.123 & 0.052 & 0.051 & 0.016 & 0.012 & 0.063 & 0.069 & 0.045 & 0.030 & 0.014 &0.007\\
      MSE      & 0.116 & 0.210 & 0.067 & 0.078 & 0.017 & 0.014 & 0.088 & 0.107 & 0.058 & 0.040 & 0.015 &0.009\\
    \hline\hline
    \end{tabular}}
    \caption{Mean, bias, variance and MSE of $\hat\theta_1, \hat\theta_2$ for bivariate exponential regression}\label{tab2}
   \end{table}

   \begin{table}[h!]{\footnotesize
     \begin{tabular}{c|cccccc|cccccc}
     \hline\hline
     &\multicolumn{6}{c}{n=500} \vline & \multicolumn{6}{c}{n=800} \\ \hline
     $\sigma_u^2$ &\multicolumn{2}{c}{0.25}& \multicolumn{2}{c}{0.2} &
     \multicolumn{2}{c}{0.1}\vline & \multicolumn{2}{c}{0.25}&
     \multicolumn{2}{c}{0.2} & \multicolumn{2}{c}{0.1}\\ \hline
      Mean     & 0.614 & 1.158 & 0.577 & 1.085 & 0.531 & 1.025 & 0.592 & 1.094 & 0.564 & 1.057 & 0.516 & 1.024\\
      Bias     & 0.114 & 0.158 & 0.077 & 0.085 & 0.031 & 0.025 & 0.092 & 0.094 & 0.064 & 0.057 & 0.016 & 0.024\\
      Variance & 0.047 & 0.054 & 0.031 & 0.019 & 0.010 & 0.006 & 0.037 & 0.024 & 0.021 & 0.012 & 0.006 & 0.004\\
      MSE      & 0.060 & 0.079 & 0.037 & 0.026 & 0.011 & 0.006 & 0.045 & 0.033 & 0.025 & 0.015 & 0.006 & 0.004\\
    \hline\hline
    \end{tabular}}
    \caption{Mean, bias, variance and MSE of $\hat\theta_1, \hat\theta_2$ for bivariate exponential regression}\label{tab3}
  \end{table}
  From Table \ref{tab2} and \ref{tab3}, we can see that the patterns for the bivariate case are similar to the univariate case as in Table \ref{tab1}.\vskip 0.1in

  \noindent{\it Quantile Regression.} In this simulation study, we consider a quantile regression model with univariate predictor, as discussed in Example 6 in Section 2. The simulated data are generated from the linear regression model $Y=\beta_0+\beta_1 X+\vep$, where $X$ has a standard normal distribution. For $\vep$, we consider two distributions -- the standard normal distribution and $(\chi^2_2-1.3863)/2$, where $\chi^2_2$ denotes a $\chi^2$-distribution with degrees of freedom $2$. Note that $1.3863$ and $2$ are the $50$-th percentile and the standard deviation of the $\chi^2$-distribution, respectively, so $\vep$ has a median of $0$ and variance $1$. In the simulation study, the sample size is chosen to be $n=300$, and $n=500$, and two measurement error variances, $0.1^2$ and $0.5^2$, are used to evaluate the effect of the measurement error on the quantile regression line.

    \begin{figure}[h!]
      \centering
      \includegraphics[width=5.5in,height=2.5in]{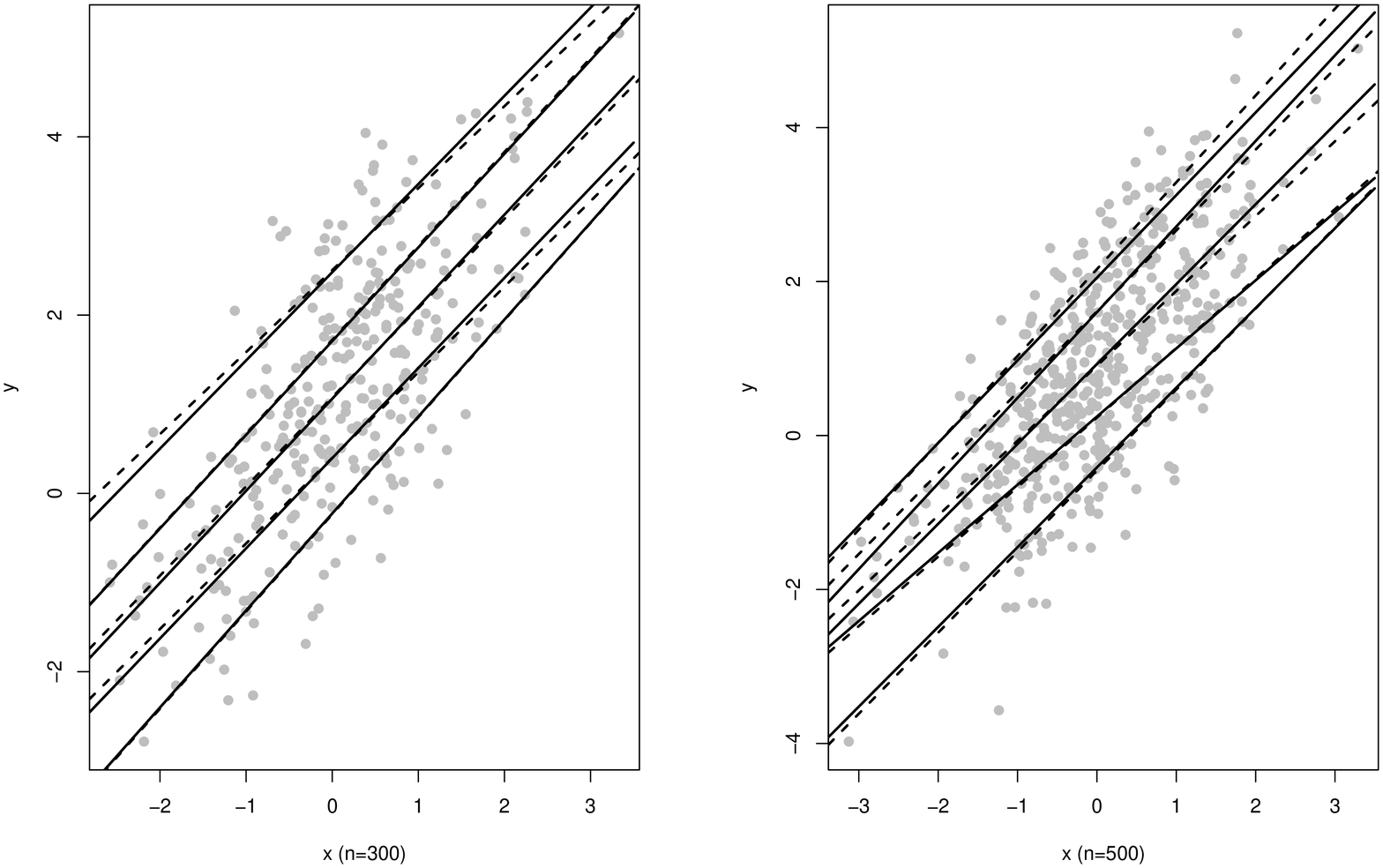}
      \caption{Quantile Regression (Normal, $\sigma_u=0.1$)}
      \label{fig2}
    \end{figure}

\begin{figure}[h!]
      \centering
      \includegraphics[width=5.5in,height=2.5in]{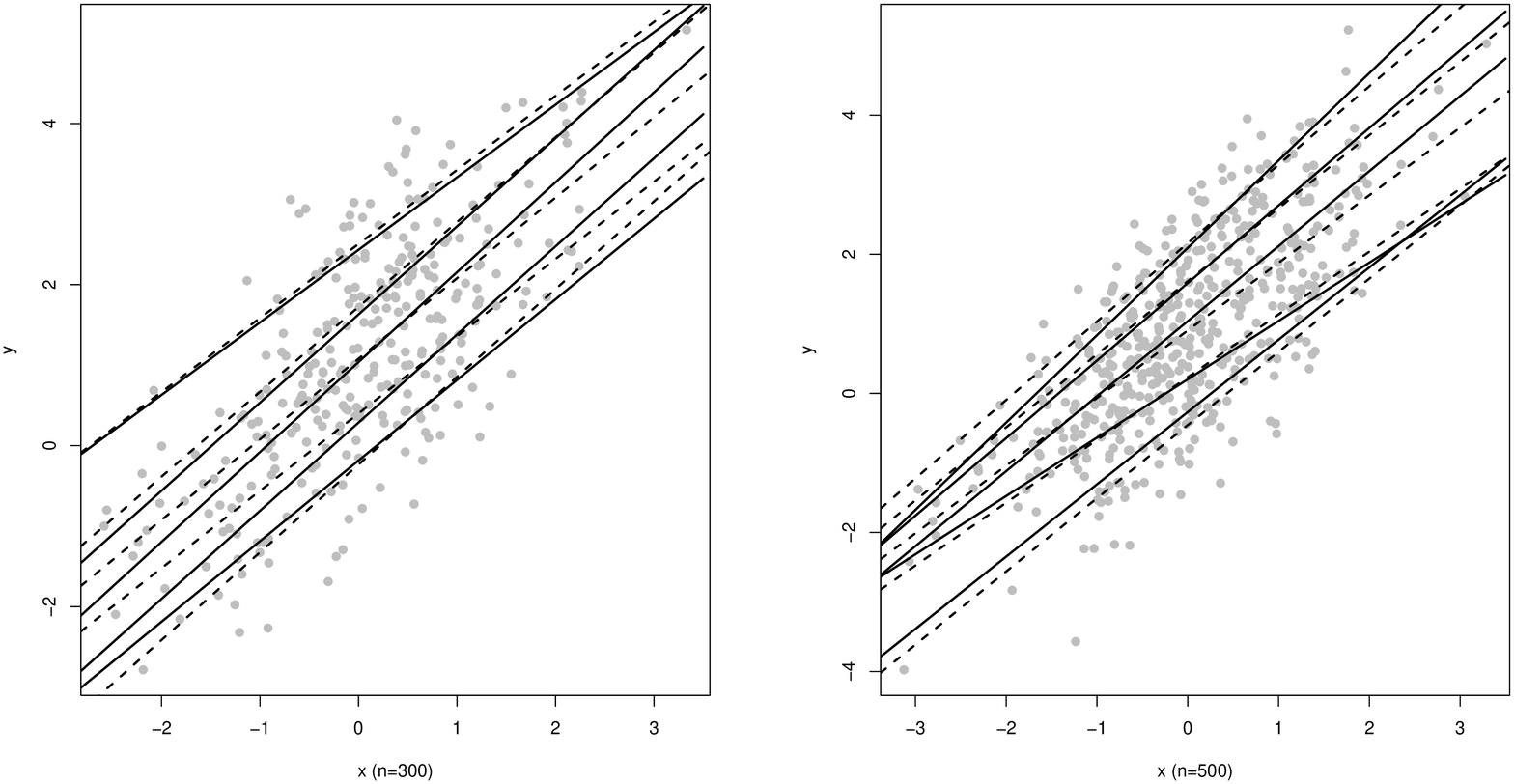}
      \caption{Quantile Regression (Normal, $\sigma_u=0.5$)}
      \label{fig3}
    \end{figure}

  The left plot on Figure \ref{fig2} shows the fitted quantile regression lines using the proposed method when $n=300$ and $\sigma_u=0.1$, when $\vep$ is normally distributed. From the top to the bottom, the solid lines are corresponding to $\tau=0.90, 0.75, 0.50, 0.25$ and $0.1$. For comparison purpose, we also plot the quantile regression lines fitted from the data on $(Y,X)$. Clearly, these two sets of quantile lines are well-matched. The right plot on Figure \ref{fig2} is for $n=500$, and the proposed method performed very well too. Figure \ref{fig3} shows the simulated quantile regression lines for $\sigma_u=0.5$. Compared to Figure \ref{fig2}, the fitted quantile regression lines deviate from the ones based on data from $(Y,X)$, but they are pretty well aligned.

\begin{figure}[h!]
      \centering
      \includegraphics[width=5.5in,height=2.5in]{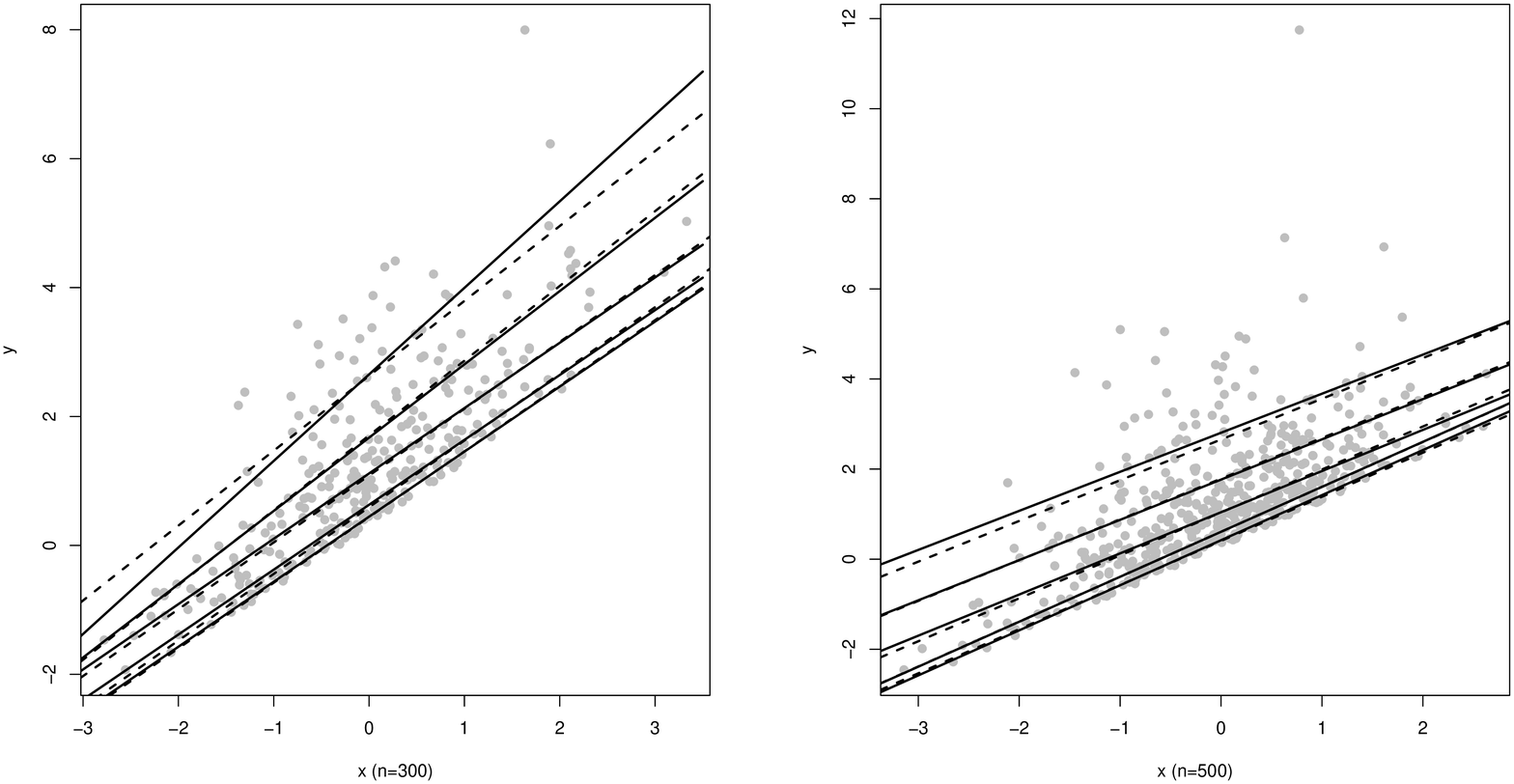}
      \caption{Quantile Regression ($\chi^2$, $\sigma_u=0.1$)}
      \label{fig4}
    \end{figure}

\begin{figure}[h!]
      \centering
      \includegraphics[width=5.5in,height=2.5in]{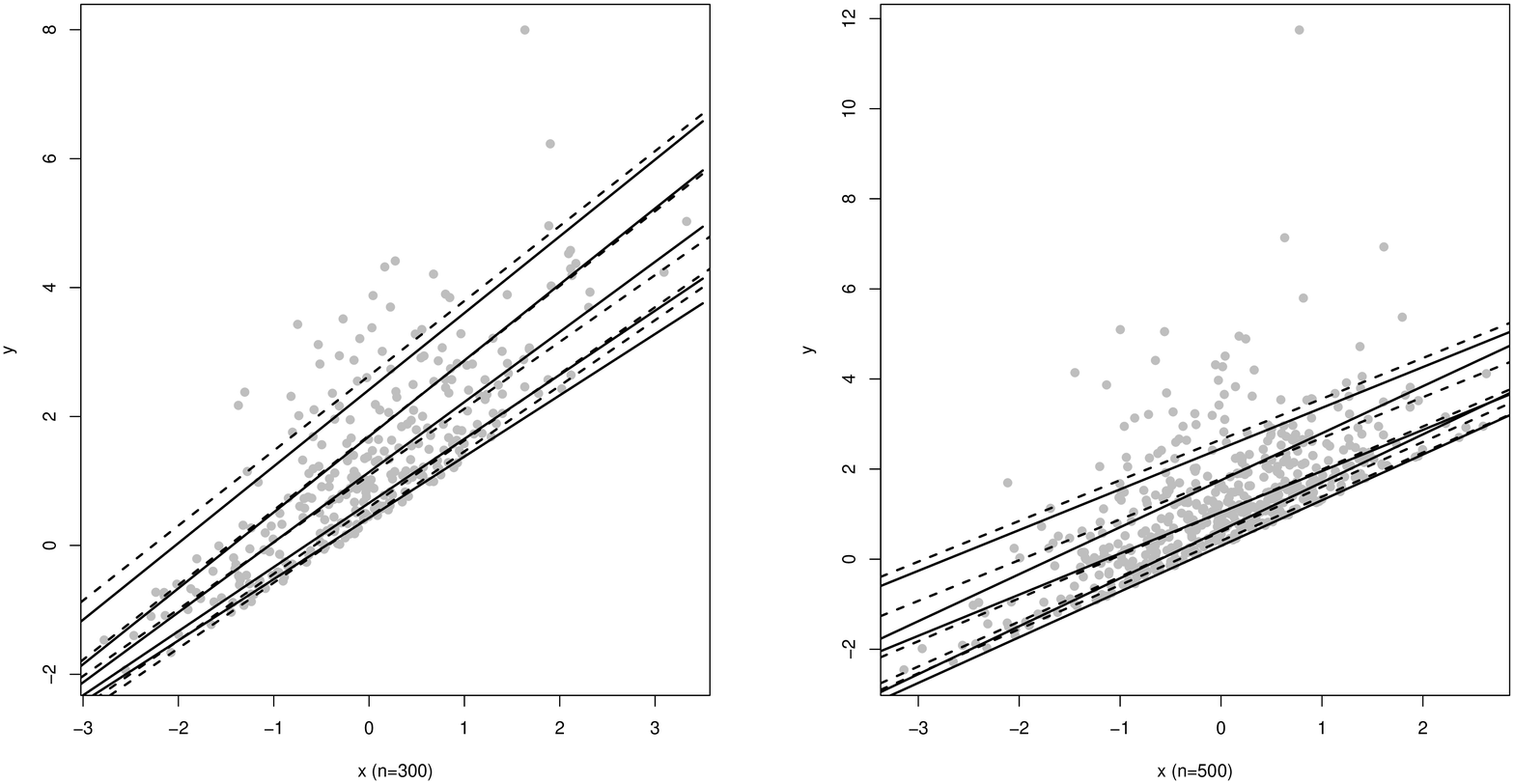}
      \caption{Quantile Regression ($\chi^2$, $\sigma_u=0.5$)}
      \label{fig5}
    \end{figure}

  The results from the simulation studies when $\vep$ has the same distribution as $(\chi^2_2-1.3863)/2$ are shown in Figures \ref{fig4} and \ref{fig5}. Note that in all scenarios, it seems like the proposed method is not very satisfying when estimating the $90$-th quantile line. This might be partially due to the fact that a right-skewed error distribution tends to generate more large outliers.

\begin{figure}[h!]
      \centering
      \includegraphics[width=5.5in,height=2.5in]{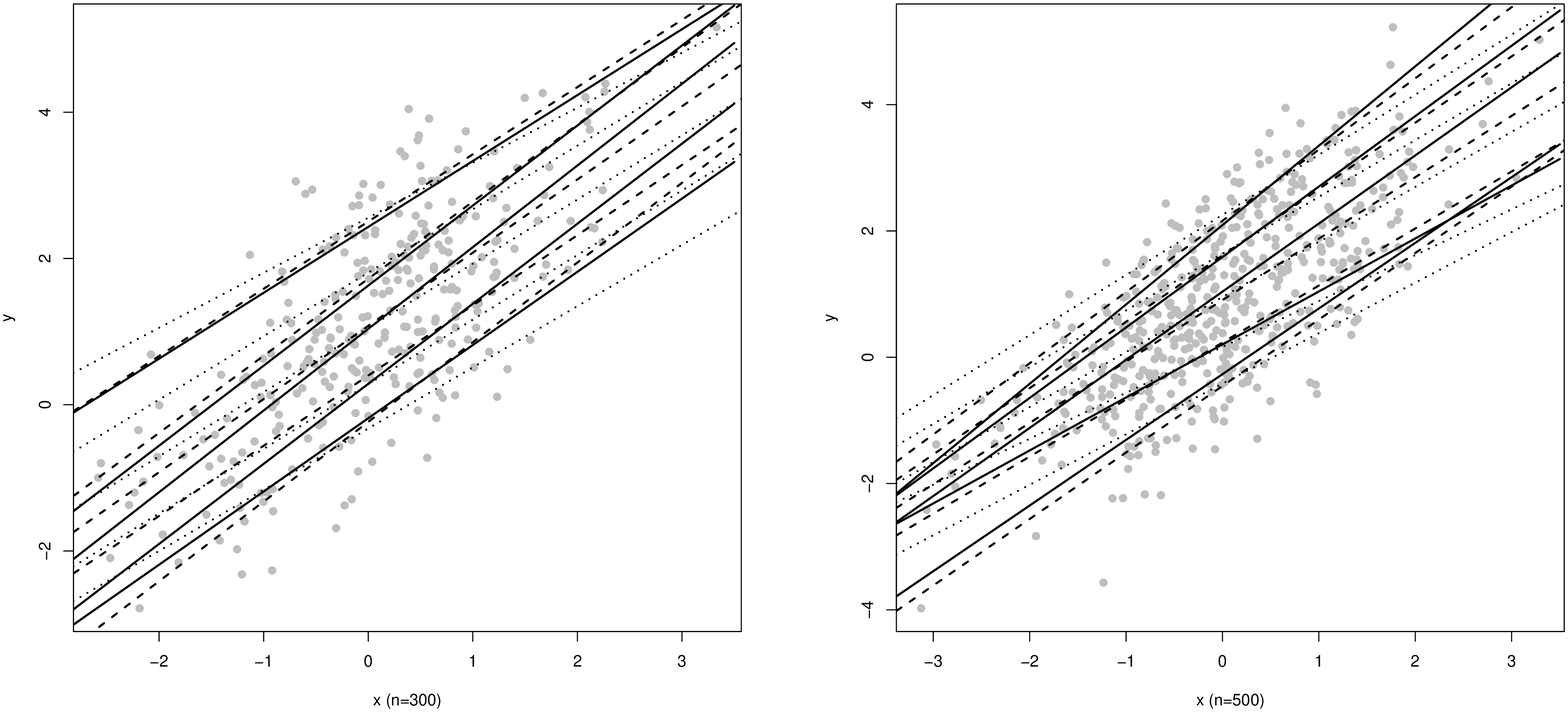}
      \caption{Naive Quantile Regression (Normal, $\sigma_u=0.5$)}
      \label{fig6}
    \end{figure}

 Finally, when $\vep$ follows a standard normal distribution, we also conduct a simulation study to compare the relative performance of the proposed method and the naive quantile regression by directly replacing $X$ with $Z$ in the standard quantile regression analysis. In Figure \ref{fig6}, the solid lines are fitted from the proposed method, the dashed lines are fitted from standard quantile regression using data on $(Y,X)$, and the dotted lines are fitted from standard quantile regression using data on $(Y,Z)$. It is clearly seen that the proposed method works much better than the naive one in reducing the bias of the estimated quantile regression coefficients.

 \subsection{Real Data Application}

 To investigate whether the serum 25-hydroxyvitamin D (25(OH)D) is influenced by the long term vitamin D average intake or not, \cite{brenna2017} analyzed a data set from the National Health and Nutrition Examination Survey (NHANES), and used a nonlinear function to model the regression mean of 25(OH)D on the long term vitamin D average intake. In this section, we apply the proposed estimation procedure on a subset of the 2009-2010 NHANES study. The selected data set contains dietary records of $806$ Mexican American females. The long term vitamin $D$ average intake $X$ is not measured directly, instead, two independent daily observations of vitamin D intake are collected. Let $Z_{ji}$ be the vitamin D intake from the $i$-th subject at the $j$-th time, and we assume that the additive structures $Z_{ji}=X_i+U_{ji}$ hold for all $i=1,2,\ldots,806$, $j=1,2$. We use $Z_i=(Z_{1i}+Z_{2i})/2$ to represent the observed vitamin intake, and by assuming that $U_{1i}$ and $U_{2i}$ are independently and identically distributed, we can estimate the standard deviation of the measurement error $U$ by the sample standard deviation of the differences $(Z_{1i}-Z_{2i})/2$, $i=1,2,\ldots,n$. Similar to \cite{brenna2017}, a square root transformation of the 25(OH)D is used to achieve a more symmetric structure.

\begin{figure}
      \includegraphics[width=5.5in,height=4in]{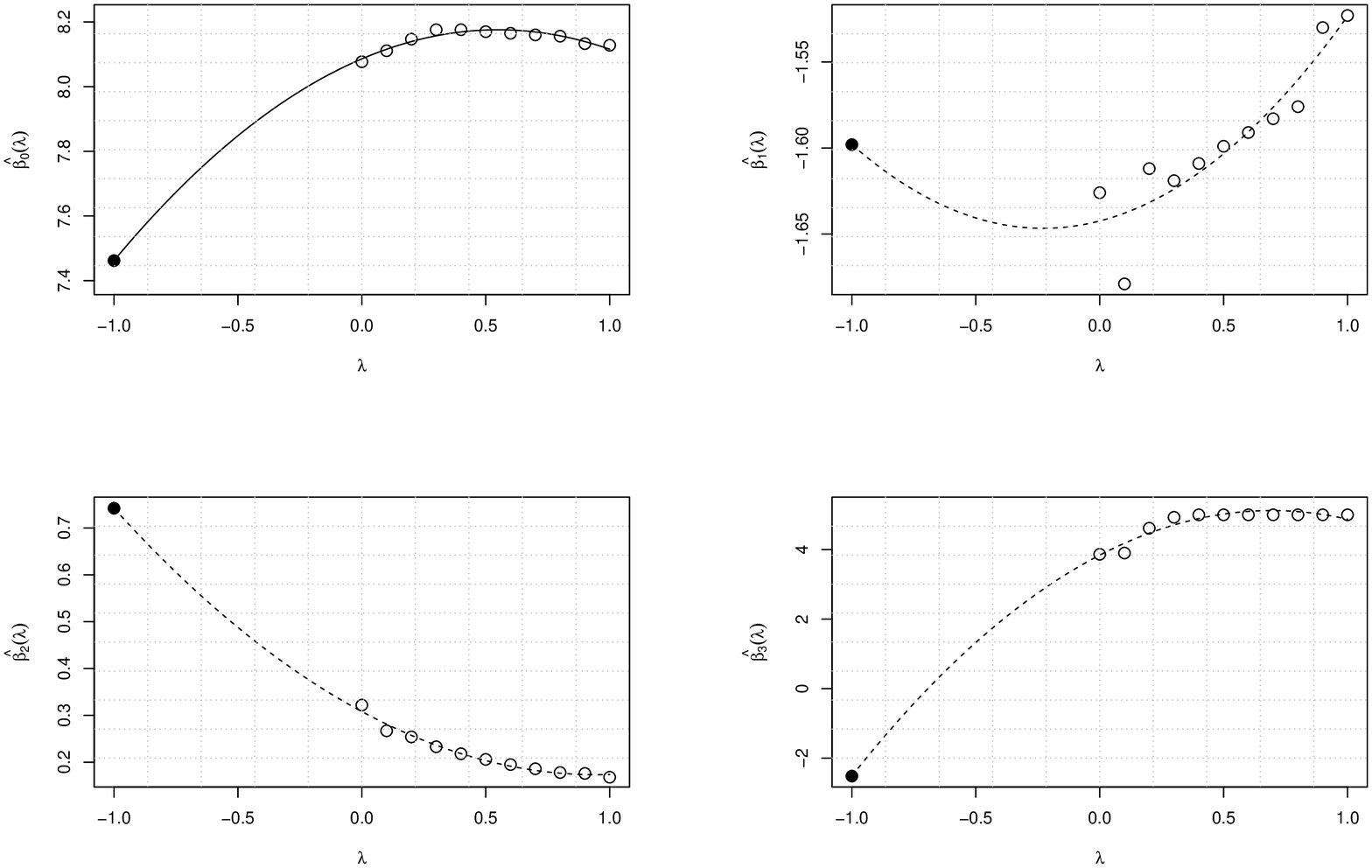}
      \caption{Extrapolation plots using quadratic function}
      \label{fig7}
    \end{figure}

\begin{figure}
      \includegraphics[width=3.5in,height=3.2in]{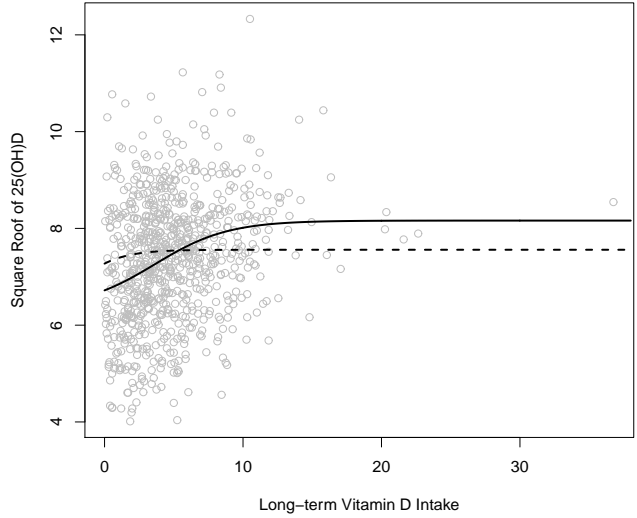}
      \caption{Nonlinear regression for HNANES data. The solid line represents the fit from the quadratic
               extrapolation and the dashed line represents the fit from the linear extrapolation}
      \label{fig8}
    \end{figure}

 We adopt the S-shaped function $\beta_0+\beta_1/(1+\exp\{\beta_2(X-\beta_3)\})$ as the regression function of $Y$ against $X$, which is also used in \cite{brenna2017}. 11 equally spaced values are chosen from $[0,1]$ as the $\lambda$ values, and both linear and quadratic extrapolation functions are tried to obtain the extrapolation estimates of the unknown parameters. The naive estimate are $\beta_0=8.077$, $\beta_1=-1.626$, $\beta_2=0.322$, $\beta_3=3.864$, and the extrapolation estimates are $\beta_0=7.462$, $\beta_1=-1.598$, $\beta_2=0.742$, $\beta_3=-2.515$. Note that the estimates for $\beta_3$ from these two methods have different signs. Figure \ref{fig7} illustrates the extrapolation processes using the quadratic function. After looking at the extrapolation plot for $\beta_3$, it is easy to see that the quadratic extrapolation function is more proper. The solid line in Figure \ref{fig8} represents the fitted regression function from the quadratic extrapolation procedure. To compare, we also plot the fitted regression using the linear extrapolation function. Clearly, the fitted regression function based on the quadratic extrapolation function seems to fit the data structure better than the one based on the linear extrapolation function.

  \section{Discussion}\label{sec5}

   The extrapolation estimation procedure proposed in this paper has a potential extension to other more complicated regression models when some predictors are contaminated with normal measurement errors, such as the partially linear regression models, the varying coefficients regression models or other semi-parametric models, where the error-prone variables appear as a linear component.

   The extension to the partially linear regression model $Y=X^T\beta+g(T)+\vep$ is straightforward, where $X$ is a $p$-dimensional latent predictor, $g$ is an unknown univariate function satisfying some smoothness conditions. Suppose $Z=X+U$, $U\sim N(0,\Sigma_u)$, and $T$ can be observed directly. Then, using the least squares procedure, and following the protocol of the extrapolation algorithm described in Section \ref{sec3}, we can estimate $\beta$ and $g$ by minimizing the conditional expectation $E\left[\sum_{i=1}^n[Y_i-Z_i^T(\lambda)\beta-g(T_i)]^2|\bY,\bZ,\bT\right]$
   or $\sum_{i=1}^n[Y_i-Z_i^T\beta-g(T_i)]^2+\lambda\beta^T\Sigma_u\beta$,
   where $\bY=(Y_1,\ldots, Y_n)^T$, $\bZ=(Z_1,\ldots, Z_n)^T$, and $\bT=(T_1,\ldots, T_n)^T$.
   Note that $\lambda=-1$ can be directly plugged in. Applying the profile least squares procedure, that is replacing $g(T_i)$ by its pseudo-kernel estimate $\hat g(t;\beta)=\sum_{j=1}^nL_h(t-T_j)(Y_j-Z_j^T\beta)/\sum_{j=1}^nL_h(t-T_j)$, where $L_h(\cdot)=L(\cdot/h)/h$ for a kernel function $L$ and a bandwidth $h$, we can estimate $\beta$ by the minimizer of $\sum_{i=1}^n[Y_i-Z_i^T\beta-\hat g(T_i)]^2-\beta^T\Sigma_u\beta$ and eventually estimate $g(t)$ with $\hat g(t; \hat\beta)$. This is the same estimation procedure used in \cite{liang1999} when they discuss the variable selection for the partially linear models with measurement errors.


   The varying coefficient regression model assumes that a scalar response variable $Y$ depends on the explanatory vectors $X$ and $W$ via the relationship $Y=X^Tg(W^T\beta)+\vep$, where $X$ is a latent vector following the measurement error structure $Z=X+U$ with $U\sim N(0,\Sigma_u)$, and $W$ is observable. The unknown vector function $g$ and the parameter $\beta$ are the quantities of interest. Pretending $g$ is known, then we can estimate $\beta$ by minimizing the conditional expectation $ E\left[\sum_{i=1}^n[Y_i-Z_i^T(\lambda)g(W_i^T\beta)]^2|\bY,\bZ,\bW\right]$ or $ \sum_{i=1}^n[Y_i-Z_i^Tg(W_i^T\beta)]^2+\lambda g^T(W_i^T\beta)\Sigma_ug(W_i^T\beta)$. Similar to the partially linear regression case, one can apply the two-step profile least squares estimation procedure to estimate $g$ and $\beta$ after replacing $\lambda$ with $-1$.

   The extension to purely nonparametric regression models is more challenging. All such possible extensions deserve independent evaluations and studies.

   Another important question is the robustness of proposed extrapolation procedure against the misspecification of the normality assumption on the measurement error. Many researchers have claimed that the classcial SIMEX procedure is robust. However, \cite{song2014} proved theoretically, along with some examples, that this statement is not true. In fact, we can see that the procedure proposed in this paper is not robust either. To see this, let us revisit the Poisson regression model discussed in Section \ref{sec2} with a univariate predictor $X$. Suppose the measurement error $U$ is normally distributed and the true parameter value is $\theta_0$, then by adding extra normal error, we have $ E[YZ\theta-\exp(Z\theta)\exp(\lambda\theta^2\sigma_u^2/2)]$ equals
      $$
       E[YX\theta-\exp(X\theta)\exp((\lambda+1)\theta^2\sigma_u^2/2)]
       =E[X\theta \exp(X\theta_0)-\exp(X\theta)\exp((\lambda+1)\theta^2\sigma_u^2/2)].
      $$
   Extrapolating $\lambda\to -1$, one can see that $\theta_0$ is a maximizer of $E[X\theta \exp(X\theta_0)-\exp(X\theta)]$. But if $U$ has a Laplace distribution with mean $0$ and variance $\sigma_u^2$, and we still proceed by adding extra normal errors, then, note that for such a Laplace random variable $U$, we have $E\exp(U\theta)=(1-\sigma_u^2\theta^2)^{-1}$, so
    $$
       E[YZ\theta-\exp(Z\theta)\exp(\lambda\theta^2\sigma_u^2/2)]=
       E[YX\theta-\exp(X\theta)(1-\sigma_u^2\theta^2/2)^{-1}\exp(\lambda\theta^2\sigma_u^2/2)].
    $$
   If we also assume that $X\sim N(0,\sigma_u^2)$, then extrapolating $\lambda\to -1$, the above expectation becomes
   $\theta\theta_0\sigma_u^2\exp(\theta_0^2\sigma_u^2/2)-(1-\sigma_u^2\theta^2/2)^{-1}$, and its maximum does not achieve at $\theta=\theta_0$ in general.

 \section{Appendix}\label{appendix}

   \noindent{\it Proof of Theorem \ref{thm1}:} Under (C1) through (C3), from Lemma 2 of \cite{jenrich1969}, we know that for every $\lambda\geq 0$, there exists a measurable function $\hat\theta_n$ such that $\hat\theta_n(\lambda)=\mbox{argmin}_{\theta\in\Theta}L_n(\theta;\lambda)$. Moreover, based on Theorem 16(a) in \cite{ferg1996}, we can also show that,
        $ \sup_{\theta\in\Theta}\left|L_n(\theta;\lambda)-L(\theta;\lambda)\right|\to 0$
  with probability $1$. Finally, by (C4), we can show that
     $ \hat\theta_n(\lambda)\to\theta(\lambda)$ in probability.
  To see the approximation of $\theta(\lambda)$ to $\theta_0$ as $\lambda\to -1$, we note that
    \begin{eqarray*}
      L(\theta; \lambda)&=&E\int [Y-m(Z+u;\theta)]^2\phi(u,0,\lambda\Sigma_u)\\
                        &=& E\int \left\{\vep^2+[m(X;\theta_0)-m(X+u;\theta)]^2\right\}\phi(u,0,(\lambda+1)\Sigma_u)du\\
                        &=& E\int \left\{\vep^2+[m(X;\theta_0)-m(X+\sqrt{(\lambda+1)}\Sigma_u^{1/2}u;\theta)]^2\right\}\phi(u,0,I)du,
    \end{eqarray*}\\ \vskip -0.5in \noindent
  it is easy to see that as $\lambda\to -1$,
    $$
      \sup_{\theta\in\Theta}\left|L(\theta; \lambda)-\sigma^2_\vep-\int [m(x;\theta)-m(x;\theta_0)]^2f_X(x)dx\right|\to 0.
    $$
  Then from {\bf (C4)}, we have $\theta(\lambda)\to\theta_0$ as $\lambda\to -1$.

  Therefore, the derivative $\dot L(\theta; \lambda)$ equals
    \begin{eqarray*}
       \iint [m(x+\sqrt{(\lambda+1)}\Sigma_u^{1/2}u;\theta)-m(x;\theta_0)]\dot m(x+\sqrt{(\lambda+1)}\Sigma_u^{1/2}u;\theta)\phi(u,0,I)f_X(x)dudx.
    \end{eqarray*}\\ \vskip -0.4in \noindent
  By Taylor expansion, we have
    $$
     m(x+\sqrt{(\lambda+1)}\Sigma_u^{1/2}u;\theta)=m(x;\theta)+\sqrt{(\lambda+1)} u^T\Sigma_u^{1/2} m'(x;\theta)
     +\frac{1}{2}(\lambda+1)u'\Sigma_u^{1/2}m''(\tilde x;\theta)\Sigma_u^{1/2}u,
    $$
  and
    \begin{eqarray*}
    \dot m(x+\sqrt{(\lambda+1)}\Sigma_u^{1/2}u;\theta)&=&\dot m(x;\theta)+\sqrt{(\lambda+1)} \dot m'(x;\theta)\Sigma_u^{1/2}u\\
    &&+\frac{1}{2}(\lambda+1)(I_{q\times q}\otimes u'\Sigma_u^{1/2})\mbox{diag}\left(
    \dot m_j''(x^*;\theta)\right)(I_{q\times q}\otimes \Sigma_u^{1/2}u),
    \end{eqarray*}\\ \vskip -0.5in \noindent
  where $\dot m_j''(x;\theta)=\partial m''(x;\theta)/\partial\theta_j$, $j=1,2,\ldots,q.$
  Therefore,
   \begin{eqarray*}
     0&=&\dot L(\theta(\lambda); \lambda)\\
      &=&\int [m(x;\theta(\lambda))-m(x;\theta_0)]\dot m(x;\theta(\lambda))f_X(x)dx\\
      & &+\frac{1}{2}(\lambda+1)\iint [m(x;\theta(\lambda))-m(x;\theta_0)](I_{q\times q}\otimes u'\Sigma_u^{1/2})\mbox{diag}\left(
    \dot m_j''(x^*;\theta)\right)\\
      && \hskip 0.2in (I_{q\times q}\otimes \Sigma_u^{1/2}u)f_X(x)\phi(u,0,I)dxdu\\
      & &+\frac{1}{2}(\lambda+1)\iint \dot m(x;\theta(\lambda)) u^T\Sigma_u^{1/2} m''(\tilde x;\theta(\lambda))\Sigma_u^{1/2}uf_X(x)\phi(u,0,I)dxdu\\
      & &+(\lambda+1)\iint u^T\Sigma_u^{1/2} m'(x;\theta(\lambda))\dot m'(\tilde x;\theta(\lambda))\Sigma_u^{1/2}u f_X(x)\phi(u,0,I)dxdu\\
      & &+o((\lambda+1)).
   \end{eqarray*}\\ \vskip -0.5in \noindent
 Also we have
   $
     m(x;\theta(\lambda))-m(x;\theta_0)=\dot m^T(x;\tilde\theta)(\theta(\lambda)-\theta_0),
   $
 this implies that,
  \begin{eqarray*}
   &&\left[\int \dot m(x;\theta_0)\dot m^T(x;\theta_0)f_X(x)dx+o(1)\right] (\theta(\lambda)-\theta_0)\\
   &=&-\frac{1}{2}(\lambda+1)\iint \dot m(x;\theta_0) u^T\Sigma_u^{1/2} m''(x;\theta_0)\Sigma_u^{1/2}uf_X(x)\phi(u,0,I)dxdu\\
      & &-(\lambda+1)\iint u^T\Sigma_u^{1/2}m'(x;\theta_0)\dot m'(x;\theta_0)\Sigma_u^{1/2}u f_X(x)\phi(u,0,I)dxdu+o((\lambda+1))\\
   &=&-\frac{1}{2}(\lambda+1)\int \dot m(x;\theta_0) \mbox{trace}(m''(x;\theta_0)\Sigma_u^{2})f_X(x)dx\\
      & &-(\lambda+1)\int \dot m'(x;\theta_0)\Sigma_u^2 m'(x;\theta_0) f_X(x)dx+o((\lambda+1)).
  \end{eqarray*}\\ \vskip -0.5in \noindent
 Note that $\tilde x\to x$ as $\lambda\to -1$, then we obtain the approximate expansion of $\theta(\lambda)$ as $\lambda\to -1$.
  This concludes the proof of Theorem \ref{thm1}. \hfill $\Box$ \vskip 0.2in

  \noindent{\it Proof of Theorem \ref{thm2}:} Denote
     $
        Q(y,z; \theta)=\int [y-m(z+u;\theta)]^2\phi(u,0,\lambda\Sigma_u)du.
     $
  Note that $\hat\theta_n(\lambda)$ satisfies
    $
      n^{-1}\sum_{i=1}^n\dot Q(Y_i,Z_i;\hat\theta_n(\lambda))=0.
    $
  Taylor expansion of the left hand side at $\theta=\theta(\lambda)$, the solution of
  $\dot EQ(Y,Z; \theta)=0$ leads to
    $$
    0=\frac{1}{n}\sum_{i=1}^n\dot Q(Y_i,Z_i; \theta(\lambda))+
    \frac{1}{n}\sum_{i=1}^n\ddot Q(Y_i,Z_i;\theta_n^*(\lambda))
     \left[\hat\theta_n(\lambda)-\theta(\lambda)\right],
    $$
  where $\theta^*_n(\lambda)$ is some value between $\hat\theta_n(\lambda)$ and $\theta(\lambda)$.
  For each $\lambda>0$, note that $\hat\theta_n(\lambda)\to\theta(\lambda)$ in probability implies that $\theta^*_n(\lambda)\to\theta(\lambda)$ in probability as $n\to\infty$, so
    $$
      \frac{1}{n}\sum_{i=1}^n\ddot Q(Y_i,Z_i;\theta_n^*(\lambda))\to \Sigma_1(\lambda)
    $$
  in probability as $n\to\infty$, where
   \begin{eqarray*}
     \Sigma_1(\lambda)&=&E\int \dot m(Z+u;\theta(\lambda))\dot m^T(Z+u;\theta(\lambda))\phi(u,0,\lambda\Sigma_u)du\\
       &&-E\int [Y-m(Z+u,\theta(\lambda))]\ddot m(Z+u;\theta(\lambda))\phi(u,0,\lambda\Sigma_u)du.
    \end{eqarray*}\\ \vskip -0.5in \noindent

  The asymptotic joint normality of $\hat\theta(\Lambda)$ is an application of the multivariate central limit theorem on $n^{-1/2}\sum_{i=1}^n\dot Q(Y_i,Z_i; \theta(\lambda_j)), j=1,2,\ldots,K$. It is sufficient to check the covariance matrix $\Sigma_0(\lambda_j, \lambda_l)$ of $\dot Q(Y, Z; \theta(\lambda_j))$ and $\dot Q(Y, Z; \theta(\lambda_l))$ has the specified form for $j, l=1,2,\ldots,K$. To see this, we need a general result for multivariate normal density functions. Without loss of generality, let $j=1, l=2$. Denote $\phi(x; \mu_1, \Sigma_1)$ and $\phi(x; \mu_2, \Sigma_2)$ as two $p$-dimensional normal density functions. Then
   \begin{equation}\label{eq6.1}
     \phi(x; \mu_1, \Sigma_1)\phi(x; \mu_2,\Sigma_2)=\phi(\mu_1; \mu_2, \Sigma_1+\Sigma_2)\phi(x;\mu,\Sigma),
   \end{equation}
 where  $\Sigma=(\Sigma_1^{-1}+\Sigma_2^{-1})^{-1}, \mu=\Sigma(\Sigma_1^{-1}\mu_1+\Sigma_2^{-1}\mu_2)$.
 Thus the covariance matrix of $\dot Q(Y, Z; \theta(\lambda_1))$ and $\dot Q(Y, Z; \theta(\lambda_2))$ is
    \begin{eqarray*}
      &&E\iint [Y-m(Z+u,\theta(\lambda_1))][Y-m(Z+v,\theta(\lambda_2))]\dot m(Z+u;\theta(\lambda_1))\dot m^T(Z+v;\theta(\lambda_2))\\
      &&\hskip 0.5in \phi(u,0,\lambda_1\Sigma_u)\phi(v,0,\lambda_2\Sigma_u)dudv.
    \end{eqarray*}\\ \vskip -0.5in \noindent
 By changing variables, the above expectation equals
    \begin{align*}
      &E\iiint [Y-m(X+w+u,\theta(\lambda_1))][Y-m(X+w+v,\theta(\lambda_2))]\dot m(X+w+u;\theta(\lambda_1)) \\
      &\hskip 0.5in \dot m^T(X+w+v;\theta(\lambda_2))\phi(u,0,\lambda_1\Sigma_u)\phi(v,0,\lambda_2\Sigma_u)\phi(w,0,\Sigma_u)dwdudv\\
      =&E\iiint [Y-m(X+u,\theta(\lambda_1))][Y-m(X+v,\theta(\lambda_2))]\dot m(X+u;\theta(\lambda_1)) \\
      &\hskip 0.5in \dot m^T(X+v;\theta(\lambda_2))\phi(w,u,\lambda_1\Sigma_u)\phi(w,v,\lambda_2\Sigma_u)\phi(w,0,\Sigma_u)dwdudv.
    \end{align*}
 From (\ref{eq6.1}), we can write $\phi(w,u,\lambda_1\Sigma_u)\phi(w,v,\lambda_2\Sigma_u)\phi(w,0,\Sigma_u)$ as either
   $$
     \phi(u,v,(\lambda_1+\lambda_2)\Sigma_u)\phi\left(\frac{\lambda_1v+\lambda_2u}{\lambda_1+\lambda_2},0,
     \frac{\lambda_{12}}{\lambda_1+\lambda_2}\Sigma_u\right)
     \phi\left(w,\frac{\lambda_1v+\lambda_2u}{\lambda_{12}},\frac{\lambda_1\lambda_2}{\lambda_{12}}\Sigma_u\right)
   $$
 or
   $$
     \phi(v,0,(\lambda_2+1)\Sigma_u)\phi\left(u,\frac{v}{\lambda_2+1},\frac{\lambda_{12}}{\lambda_2+1}
     \Sigma_u\right)
     \phi\left(w,\frac{\lambda_1v+\lambda_2u}{\lambda_{12}},\frac{\lambda_1\lambda_2}{\lambda_{12}}\Sigma_u\right),
   $$
 where $\lambda_{12}=\lambda_1+\lambda_2+\lambda_1\lambda_2$.
 This, together with $Y=m(X;\theta_0)+\vep$, implies that the covariance matrix of $\dot Q(Y, Z; \theta(\lambda_1))$ and $\dot Q(Y, Z; \theta(\lambda_2))$ can be written as $\Sigma_0(\lambda_1, \lambda_2)$ as defined in Theorem \ref{thm2}.
\hfill $\Box$\vskip 0.2in


\noindent{\it Proof of Theorem \ref{thm3}:}
 Note that $\hat\Gamma$ is the minimizer of (\ref{eq3.4}), by a Taylor expansion, we have
 \begin{eqarray*}
    &&0=\dot G^T(\Lambda,\hat\Gamma)(\hat\theta_n(\Lambda)-G(\Lambda,\hat\Gamma))\\
   &=&\dot G^T(\Lambda,\Gamma_0)(\hat\theta_n(\Lambda)-G(\Lambda,\Gamma_0))
  +\left[T(\Lambda,\tilde\Gamma)
  -\dot G^T(\Lambda,\tilde\Gamma)\dot G(\Lambda,\tilde\Gamma)
  \right](\hat\Gamma-\Gamma_0),
 \end{eqarray*}\\ \vskip -0.5in \noindent
 where $\tilde\Gamma$ is between $\hat\Gamma$ and $\Gamma_0$ and
   $$
     T(\Lambda,\Gamma)=\sum_{j=1}^M\sum_{k=1}^q \begin{pmatrix}\frac{\partial G_k(\lambda_j,\Gamma)}{\partial\gamma_1\partial\Gamma^T}(\hat\theta_{nk}(\lambda_j)-G_k(\lambda_j,\Gamma))\\
     \vdots \\
     \frac{\partial G_k(\lambda_j,\Gamma)}{\partial\gamma_d\partial\Gamma^T}(\hat\theta_{nk}(\lambda_j)-G_k(\lambda_j,\Gamma))
     \end{pmatrix}_{d\times d},
   $$
 where $\hat\theta_{nk}(\lambda)$ is the $j$-th component of $\hat\theta_n(\lambda)$, $j=1,2,\ldots,q$.
 The consistency of $\hat\Gamma$ to $\Gamma_0$ implies that
  \begin{eqarray*}
   \sqrt{n}\left[\dot G^T(\Lambda,\Gamma_0)\dot G(\Lambda,\Gamma_0)\right](\hat\Gamma-\Gamma_0)=\sqrt{n}\dot G^T(\Lambda,\Gamma_0)(\hat\theta(\Lambda)-\theta(\Lambda))+o_p(1).
  \end{eqarray*}\\ \vskip -0.5in \noindent
Recall the notations $H(\Lambda)$ and $\Pi(\Lambda)$ defined right before Theorem \ref{thm3} in Section \ref{sec3}, the asymptotic normality of $\hat\theta_n(\Lambda)$ implies that
  \begin{equation}\label{eq6.2}
    \sqrt{n}\left[\hat\Gamma-\Gamma_0\right]
    \Longrightarrow N(0,\Pi(\Lambda)).
  \end{equation}
Note that the EX estimate $\hat\theta_n$ is defined as $\hat\theta_n=G(-1,\hat\Gamma)$, also note that $G(-1,\Gamma_0)=\theta_0$, so by Taylor expansion again, $ \hat\theta_n-\theta_0=\dot G(-1, \tilde\Gamma)(\hat\Gamma-\Gamma_0)$, together with the asymptotic result (\ref{eq6.2}), we prove Theorem \ref{thm3}.\hfill $\Box$

 \bibliographystyle{elsarticle-harv}
 \bibliography{simex}

\end{document}